\documentclass[12pt]{article}
\pdfoutput=1
\usepackage[numbers,sort&compress]{natbib}
\usepackage{amssymb,amsfonts,latexsym,amsmath,amsthm,times}
\usepackage{float}
\usepackage[section]{placeins}
\usepackage{lineno}
\usepackage{graphicx}
\usepackage[english]{babel}
\usepackage{longtable,supertabular}
\usepackage{subfigure}

\numberwithin{equation}{section}

\DeclareMathOperator{\id}{id}
\begin{document}


\thispagestyle{plain}

\vspace*{2cm}

\normalsize

\begin{center}
 
{\Large \bf Competition of residents and invaders in a variable environment: Response to enemies and dangerous noise}

\vspace*{1cm}

{\bf I. Siekmann$^a$ and H. Malchow$^b$\footnote{Corresponding author. E-Mail: horst.malchow@uni-osnabrueck.de}}

\vspace*{0.5cm}

{$^a$ Systems Biology Laboratory, Melbourne School of Engineering,
  The University of Melbourne, Parkville 3010 VIC Australia}

{$^b$ Institute of Environmental Systems Research, School of
Mathematics\,/\,Computer Science, Osnabr\"uck University, Barbarastr. 12,
49076 Osnabr\"uck, Germany}

\end{center}


\vspace*{1cm}

\noindent {\bf Abstract.}
The possible control of competitive invasion by infection of the invader and multiplicative noise is studied. The basic model is the Lotka-Volterra competition system with emergent carrying capacities. Several stationary solutions of the non-infected and infected system are identified as well as parameter ranges of bistability. The latter are used for the numerical study of invasion phenomena. The diffusivities, the infection but in particular the white and colored multiplicative noise are the control parameters. It is shown that not only competition, possible infection and mobilities are important drivers of the invasive dynamics but also the noise and especially its color and the functional response of populations to the emergence of noise.

\vspace*{0.5cm}




\vspace*{1cm}

\setcounter{equation}{0}
\section{Introduction}
\label{sec:intro}

The main aim of modeling biological population dynamics is to improve the understanding of the functioning of food chains and webs as well as their dependence on internal and external conditions. Hence, mathematical models of biological population dynamics have not only to account for growth and interactions but also for spatiotemporal processes like random or directed and joint or relative motion of species, as well as the heterogeneity of the environment. Early attempts began with statistics, exponential growth, physico-chemical (neutral) diffusion, and Lotka-Volterra type interactions. These approaches have been continuously refined to more realistic descriptions of the development of natural populations.\\
Ecological and epidemiological models are known since more than 200 years. First attempts to merge these models appeared only about 30 years ago, cf. \cite{And:86, Had:89, Fre:90a, Gao:92} as well as \cite{Ven:92,Ven:94}. Infectious diseases are prominent examples of biological invasions and continue to (re-)emerge in modern times. The negative econo-ecological effects of bioinvasions \cite{Dra:89, Pim:02} have led to a remarkable hype of bioinvasion research incl. modeling, cf. \cite{Mol:86, Hen:89, Wil:96, Shi:97, Sax:05}. The history of research on stochastic processes and integration is long as well, historical surveys have been published, cf. \cite{Jar:04, Mey:09}. The seminal work by \^Ito \cite{Ito:51} and Stratonovich \cite{Str:63} should be particularly recognized.\\
The present, to a large extent numerical study combines aspects of spatial eco-epidemiology and environmental stochasticity, namely the diffusive invasion of an alien species, its competition with the indigenous resident, and its biocontrol through targeted infection in a noisy environment. Contrary to previous publications \cite{Mal:11,Mal:13}, the environmental variability is modeled as external multiplicative noise, in some cases with a certain functional response of the populations.

  Modeling environmental variability with multiplicative white noise
  goes back to the 1970s. Not only did May \cite{May:73a,May:73d}
  introduce the model that is used until today as a perturbation of
  the growth rate of a population by ``white noise'' but only a few
  years later, a more mechanistic basis of this model was
  developed. Branching processes provide a stochastic model that
  describes the number of offspring for a given number of
  individuals~$Z_i$ within one generation~$i$. The population
  number~$Z_{i+1}$ in generation~$i+1$ is updated for a given
  population~$Z_i$ according to previously chosen probability
  distributions. This model was extended by Smith and Wilkinson
  \cite{Smi:68a, Smi:69a} by a stochastic process~$\zeta_i$ which
  modulates for each new generation the offspring that is
  generated. Branching processes in random environments~(BPRE) provide
  an individual-based model for population growth. For large
  population numbers, a BPRE can be approximated by a stochastic
  differential equation~(SDE) that accounts both for demographic as
  well as environmental stochasticity. Keiding~\cite{Kei:75a}
  conjectured the form of the resulting diffusion approximation, his
  conjecture was rigorously proven by Kurtz \cite{Kur:78b}. The model
  introduced by May is obtained from Kurtz' diffusion approximation by
  neglecting the term due to demographic stochasticity. Thus, the SDE
  model for environmental stochasticity is derived from the influence
  of a random environment on the population dynamics of a branching
  process but neglects its demographic stochasticity.

  In this study we consider the properties of environmental
  variability in more detail. The multiplicative noise term implies
  that the effect of environmental fluctuations on the individuals of
  a population is additive. Whereas this seems reasonable for small
  population densities we propose that for large population numbers
  the effect of individual responses to environmental fluctuations on
  the population should decrease. Thus, we propose that the
  population-dependent response to environmental noise saturates for
  large population numbers similar to the functional response of
  predators at large prey densities and therefore we model the
  population-dependent response to environmental noise in a completely
  analogous way.

  The subject of our study is the influence of environmental
  stochasticity on a biological invasion. In order to account for
  spatial spread we extend our system of SDEs by diffusion terms so
  that we obtain a system of stochastic reaction-diffusion
  equations. Spatiotemporal environmental fluctuations are represented
  by time-dependent random fields. In contrast to previous studies, we
  consider random fields that are correlated both in space and in time,
  i.e., spatiotemporally colored noise.

  The assumption of uncorrelated white noise is usually justified by
  the coarseness of temporal or spatial scale, respectively. If
  spatial or temporal correlation length are much shorter than the
  time or length scale of interest, it seems valid to consider the
  time-dependent random field as uncorrelated. However, particular
  care must be taken when considering spatiotemporal dynamics driven
  by noise. Stochastic differential equations driven by uncorrelated
  noise can usually solved over a function space such as~$L^2$ and
  this remains true if the system is extended to a reaction-diffusion
  equation over one-dimensional space~($d=1$). But for spatial
  dimensions~$d \geq 2$ solutions for stochastic reaction-diffusion
  equations driven by uncorrelated noise can only be guaranteed in a
  space of generalised functions, see e.g. \cite{Wal:86, Mue:09a}. The
  reason for this phenomenon is that the Laplacian cannot smooth
  uncorrelated noise sufficiently for spatial dimensions exceeding 1 so
  that a solution may contain peaks resembling the~$\delta$
  distribution. Not only is the physical significance of these
  solutions debatable but also numerical approximations are not
  capable of capturing this aspect of the continuous system. Here we
  take a pragmatic point of view on this difficult problem and present
  numerical solutions for temporally and spatially white noise where
  the space may be interpreted as a discrete lattice whose nodes
  interact by a discrete Laplacian.

\setcounter{equation}{0}
\section{Resident-invader competition with infection in the invader population}
\label{sec:cci}

For the invasion of a resident population by a competing invader, the Lotka-Volterra competition model is used, i.e.,
\begin{align}
\label{eq:resident}
\frac{dN_1}{dt} &= r_1 N_1 - c_{11} N_1^2 - c_{12} N_1 N_2,\\
\label{eq:invader}
\frac{dN_2}{dt} &= r_2 N_2 - c_{22} N_2^2 - c_{21} N_1 N_2,
\end{align}

where $N_1$ and $N_2$ are resident and invader respectively. Carrying capacities will not explicitely be introduced because they can suppress a higher variety of solutions and rather appear as special cases \cite{Ful:81, Kun:91, Mall:12, Sieb:14}. The $r$'s stand for the growth rates and the $c$'s for the inter- and intraspecific competition.\\

A specific infection of the invading population can be used as biocontrol measure to stop and reverse the invasion, cf. \cite{Har:92, Jul:97, McE:99, Coo:04}. To model this, the invader population is split into susceptibles $S$ and infecteds $I$,
\[
N_2 = S + I.
\]

Then, the local dynamics reads with notation $\mathbf{X} = \{X_1 \equiv N_1, X_2=S, X_3=I\}$

\begin{align}
\label{eq:resident1}
\frac{dX_1}{dt} &= f_1\,(\mathbf{X}) = r_1 X_1 - c_{11} X_1^2 - c_{12} X_1 (X_2 + X_3),\\
\label{eq:susinv1}
\frac{dX_2}{dt} &= f_2\,(\mathbf{X}) = r_2 X_2 - c_{22} X_2 (X_2 + X_3) - c_{21} X_1 X_2 - \lambda \frac{X_2 X_3}{(X_2+X_3)^q},\\
\label{eq:infinv1}
\frac{dX_3}{dt} &= f_3\,(\mathbf{X}) = r_2 X_3 - c_{22} X_3 (X_2 + X_3) - c_{21} X_1 X_3 + \lambda \frac{X_2 X_3}{(X_2+X_3)^q} - \mu X_3\,,
\end{align}

where $\lambda$ is the transmission coefficient of the disease and $\mu$
the disease-induced higher mortality rate of the infecteds. The exponent $q$
allows to describe mass-action type ($q = 0$) and frequency-dependent
transmission ($q = 1$) of the disease respectively \cite{And:86, McC:01}.\\

However, one cannot expect that growth rates and competition intensities of
susceptibles and infecteds are the same. They should rather be split and could
be ordered like
\begin{eqnarray}
\label{eq:split}
\begin{array}{llclc}
r_2    &\Rightarrow & \{r_2,r_3\}      &~\textrm{\&}~ & r_3 \le r_2\,, \\
c_{12} &\Rightarrow &\{c_{12},c_{13}\} &~\textrm{\&}~ & c_{13} \le c_{12}\,,\\
c_{21} &\Rightarrow &\{c_{21},c_{31}\} &~\textrm{\&}~ & c_{21} \le c_{31}\,,\\
c_{22} &\Rightarrow &\{c_{22},c_{23},c_{32},c_{33}\}  &~\textrm{\&}~ & c_{23} 
\le c_{33} \le c_{22} \le c_{32}\,.
\end{array}
\end{eqnarray}

The ordering of the intra- and interspecific competition coefficients of susceptibles and infecteds depends on
the biological species, cf. \cite{Bed:05}. However, it can be certainly accepted \cite{Sieb:14} that
\[
 c_{23} \le c_{22} \wedge c_{33} \le c_{32}\,.
\]

System (\ref{eq:resident1}--\ref{eq:infinv1}) then changes to
\begin{align}
\label{eq:resident2}
\frac{dX_1}{dt} &= f_1(\mathbf{X}) = (r_1 - c_{11} X_1) X_1 - (c_{12} X_2 + c_{13} X_3) X_1\,,\\
\label{eq:susinv2}
\frac{dX_2}{dt} &= f_2(\mathbf{X}) = (r_2 - c_{22} X_2) X_2 - (c_{21} X_1 + c_{23} X_3) X_2 -
\lambda \frac{X_2 X_3}{(X_2+X_3)^q},\\
\label{eq:infinv2}
\frac{dX_3}{dt} &= f_3(\mathbf{X}) = (r_3 - \mu - c_{33} X_3) X_3 - (c_{31} X_1 + c_{32} X_2) X_3 +
\lambda \frac{X_2 X_3}{(X_2+X_3)^q}\,.
\end{align}

For convenience, the model of the local dynamics is not analysed in terms of
$X_1$, $X_2$ and $X_3$ but rather in $X_1$, $i$ and $N_2=X_2+X_3$ where $i$ is the
prevalence, i.e., the infected fraction of the total invader population $N_2$
\cite{Hil:06b},
\[
i = \frac{X_2}{X_2+X_3} = \frac{X_2}{N_2}.
\]
Having in mind that
\begin{equation}
\frac{di}{dt} = \frac{1}{N_2} \left(\frac{dX_2}{dt} - i \frac{dN_2}{dt}\right)\,,
\end{equation}

it follows
\begin{align}
\label{eq:resident3}
\frac{dX_1}{dt} &= r_1 X_1 - c_{11} X_1^2 - \left[c_{12} (1 - i) +
c_{13} i\right] X_1 N_2\,,\\[3mm]
\label{eq:prevalence3}
\frac{di}{dt} &= \left\{r_3 - r_2 + \lambda N_2^{1-q} - \mu 
               + N_2 \left[ (c_{22}-c_{32})(1-i)+(c_{23}-c_{33})i \right] 
               + X_1 \left[c_{21} - c_{31}\right]\right\}\,*~~~~~~\nonumber\\
               &~~~~~* i (1 - i)\,,\\
\label{eq:totalinv3}
\frac{dN_2}{dt} &= \left\{ G(N_2,i) - \left[ c_{21} (1 - i) + c_{31} i \right]
X_1 \right\} N_2\,,
\end{align}

with
\begin{equation}
\label{eq:general}
G(N_2,i) = r_2 (1-i) + (r_3-\mu) i - N_2 \left[ c_{22} (1-i)^2 + (c_{23}+c_{32})
i (1-i) + c_{33} i^2 \right]\,.
\end{equation}

The latter expression is also found for predator-prey systems with infected prey
\cite{Sieb:14}. Note that if the resident $X_1$ resp. the predator in \cite{Sieb:14} cannot
distinguish between susceptible and infected invader resp. prey, the temporal
change of the prevalence becomes independent of the type of interspecific
ecological interactions such as competition and predation. It only contains
terms describing the intraspecific competition of susceptibles and infecteds in
the infected population.

\subsection{Stationary solutions and stability for frequency-dependent (standard) incidence q=1}
\label{sec:statcci}

In phytopathology, the transmission of especially fungal diseases is described with standard incidence \cite{Pla:82}. A corresponding model of the invasion of a fungal disease over a vineyard has been investigated in \cite{Bur:08a}. Further on, only the standard incidence is considered, i.e., $q = 1$.

The infection-free system, i.e., $i \equiv 0, N_2 = X_2$, is the Lotka-Volterra competition model with its known stationary solutions and their stability ranges. Especially interesting for the consideration of spatial invasions is the bistable parameter range
\[
r_1 c_{22} - r_{2} c_{12} > 0 ~\wedge~ r_2 c_{11} - r_{1} c_{21} > 0\,,
\]
when both the invader-free $(\dfrac{r_1}{c_{11}},0,0)$ and the resident-free $(0,0,\dfrac{r_2}{c_{22}})$ states are stable and can compete for space. The opposite case may also happen: The invader arrives already infected, i.e., $i \equiv 1, N_2=X_3$, and the invader-free $(\dfrac{r_1}{c_{11}},0,0)$ and the resident-free $(0,0,\dfrac{r_3-\mu}{c_{33}})$ states can be both at once stable for
\[
r_1 c_{33} - (r_{3}-\mu)\,c_{13} > 0 ~\wedge~ (r_{3}-\mu)\,c_{11} - r_{1} c_{31} > 0\,.
\]
However, the latter as well as the possible bistability of
resident-only $(\dfrac{r_1}{c_{11}},0,0)$ and
(susceptible-infected)-invader-only $(0,i^{S1},N_2^{S1})$ states will
not be considered here.

\subsection{Spatiotemporal dynamics in a variable environment}
\label{sec:ccs}

The main focus of this study is to consider the
  spatiotemporal effects of a more detailed model of environmental
  variability. We assume that all species spread randomly so that
  mobility can be described as diffusion with coefficients
  $\mathbf{D} = \{D_{ii}=D_i\,; ~D_{ij} \equiv 0 ~\forall i \ne j\,;
  ~i,j=1,2,3\}$. Also we add Gaussian random
  fields~$\boldsymbol{\xi}(\vec{r},t) = \{\xi_i(\vec{r},t)\,;
  ~i=1,2,3\}$ to system (\ref{eq:resident2}--\ref{eq:infinv2}) so that
  we obtain the system of stochastic partial differential equations
\begin{equation}
\label{eq:densvec}
\frac{\partial \mathbf{X}(\vec{r},t)}{\partial t} - \mathbf{D} \Delta \mathbf{X}(\vec{r},t) = \mathbf{f}\left[\mathbf{X}(\vec{r},t)\right] + \mathbf{g}\left[\mathbf{X}(\vec{r},t)\right]\,\boldsymbol{\xi}(\vec{r},t)\,,
\end{equation}

where the matrix function $\mathbf{g}(\mathbf{X}) = \{g_{ij}(\mathbf{X});~i,j=1,2,3\}$ determines the density-dependent noise intensity. We consider horizontal processes with position vector
$\vec{r} = \{x,y\}$ and corresponding Laplace operator
$\Delta = \partial^2/\partial x^2 + \partial^2/\partial y^2$. In
literature, often temporally and spatially uncorrelated ``white''
Gaussian fields with zero mean and delta correlation have been considered
\begin{equation}
 \left<\xi_i(\vec{r},t)\right> = 0\,,~ \left<\xi_i(\vec{r}_1,t_1)\,\xi_i(\vec{r}_2,t_2)\right> = \delta(\vec{r}_1-\vec{r}_2) \, \delta(t_1-t_2)\,,~i=1,2,3\,.
\end{equation}

Here, we investigate the effect of extending this model by correlated ``colored'' noise with correlation lenghts~$\tau$ and~$\lambda$ in the temporal and spatial domain, respectively. Apart from using colored noise we also investigate a generalisation of the density-dependent noise~$\mathbf{g}\left(\mathbf{X}\right)$. Purely diagonal, linear multiplicative noise
\begin{equation}
  \label{eq:linearnoise}
  g_{ii}(\mathbf{X}) = \omega_{ii} X_i\,; ~g_{ij}(\mathbf{X}) = 0 ~\forall ~i \ne j\,; ~i,j=1,2,3\,;
\end{equation}

can be interpreted as a model where individuals respond independently
to stochastic environmental variability. Thus, the effect of
environmental fluctuations on each individual directly translates into
variability at the population level -- the response at the population
level is additive. The alternative model suggested here is based on
the assumption that in large populations individuals do not repond to
fluctuations independently from each other. Instead we propose that
larger populations respond to environmental variability in a more
robust way, i.e., neither favourable nor adverse effects influence the
population proportional to the number of individuals:
\begin{equation}
  \label{eq:simpnoise}
   g_{ij}(\mathbf{X}) = \dfrac{\omega_{ij} X_j^m}{\gamma_{ij} + \sum\limits_{k=1}^{3}\,a_{ik}X_k^n}
    \,; ~i,j=1,2,3\,; ~1\le m\le n\le 2\,.
\end{equation}

For $m=n$, the parameter~$\omega_{ij}/a_{ij}$ is the maximum noise intensity that
is reached asymptotically for large populations~$X_j$. The
parameter~$\gamma_{ij}$ is the population level at which the noise
intensity reaches half of the maximum level~$\omega_{ij}/a_{ij}$. Thus, this
parameter describes the ability of a population to collectively reduce
the effect of noise -- the higher~$\gamma_{ij}$, the higher the population
level must be until population~$X_j$ is appreciably affected by
environmental variability.

For $m<n$, the noise intensity even decreases and eventually vanishes for high population densities. However, these values are never reached.

In previous papers \cite{Mal:11,Mal:13}, it was shown that a certain variability of the environment and the mobilities of the competitors are the system-driving forces. Extreme events such as landslides lead to bare ground re-invadable by both resident and alien species. These events at random times, size and locations are not considered here. They are replaced by white and colored noises \cite{Rip:98, Siek:09, Siek:13}. Again, the biocontrol of the invasion through a specific infection of the non-indigenous species is studied.

\setcounter{equation}{0}
\section{Numerical methods}
\label{sec:numerical}

The numerical solution of stochastic partial differential equations is
a difficult problem and the subject of current research. For this
reason we explain how the spatiotemporal model~\eqref{eq:densvec} can
be solved numerically and how spatiotemporally correlated noise can be
generated. We follow a finite difference approach where in a first step the
spatial domain is discretised. In this way the system of stochastic
partial differential equation is approximated by uncoupled stochastic
differential equations that are solved numerically in a second
step. For the first step, we use the semi-implicit Peaceman-Rachford
method \cite{Pea:55,Tho:95} which, in particular for stochastic equations, often seems to
be more robust than the simplest explicit scheme. In the following we
will explain how the discretised system can be solved using the
derivative-free Milstein method and how spatiotemporally correlated
noise can be generated.

\subsection{Derivative-free Milstein method}
\label{sec:milstein}

For numerical integration, the derivative-free Milstein method is used, cf. \cite{Mil:88,Mil:95,Klo:99} but also the short descriptions \cite{Schaf:10, Gar:11}. Sometimes and in particular for the purpose of this study, it is even sufficient to consider purely diagonal intensity matrices
\begin{equation}
  \label{eq:simpnoise1}
   g_{ii}(\mathbf{X}) = \dfrac{\omega_{ii} X_i^m}{\gamma_{ii} + a_{ii}X_i^n}\,; ~a_{ii}\equiv 1\,; ~i=1,2,3\,; ~1\le m\le n\le 2\,.
\end{equation}

Then, the Milstein scheme reads with time step $\Delta t$ and Stratonovich interpretation
\begin{align}
 X_{t+\Delta t}^i &= X_t^i + f_i(X_t^i) \Delta t + g_{ii}(X_t^i) \Delta W_i + \dfrac{1}{2\sqrt{\Delta t}}\left[g_{ii}(\bar{X}_t^i) -g_{ii}(X_t^i\right] (\Delta W_i)^2\,,\\
 \text{with}&\nonumber\\
 \bar{X}_t^i &= X_t^i + f_i(X_t^i) \Delta t + g_{ii}(X_t^i) \sqrt{\Delta t}\,,\nonumber\\
 \text{and}&\nonumber\\
 \Delta W_i &= W_{t+\Delta t}^i - W_t^i \sim \sqrt{\Delta t} \, \mathcal{N}(0,1)\,.\nonumber
\end{align}
As usual, $\mathcal{N}(0,1)$ stands for the normal distribution with zero mean and unity variance. The required uniformly distributed random numbers are generated with the Mersenne Twister \cite{Mat:98b}, the normally distributed with the common Box-Muller algorithm \cite{Box:58}.

\subsection{Generation of correlated Gaussian random fields}
\label{sec:colorednoise}

Garc{\'i}a-Ojalvo and Sancho \cite{Gar:92a,Gar:99} developed a method for
generating spatially and temporally colored noise from the stochastic
reaction-diffusion equation
\begin{equation}
  \label{eq:coloredspattemp}
  \frac{\partial \zeta(\vec{r},t)}{\partial t} = \frac{\lambda^2}{\tau} \Delta \zeta(\vec{r},t) -\frac{1}{\tau}\zeta(\vec{r},t) +\frac{1}{\tau} \eta(\vec{r},t)
\end{equation}

Here, the term $\eta(\vec{r},t)$ stands for uncorrelated (white) noise. The
parameters~$\tau$ and~$\lambda$ determine the correlation lengths in
the temporal and the spatial domain, respectively. In addition, a
scaling factor~$\epsilon$ for the variance of normally-distributed
random variables in Fourier space has to be chosen. The authors
introduce Eq.\,\eqref{eq:coloredspattemp} as an analogon of the
Ohrnstein-Uhlenbeck process which is the solution
of~\eqref{eq:coloredspattemp} without the spatial
term~$\dfrac{\lambda^2}{\tau} \Delta \zeta(r, t)$. But it has to be
noted that for two-dimensional space (and spatial dimensions exceeding
two) the solutions of equation~\eqref{eq:coloredspattemp} will be
generalised functions rather than functions in a space such
as~$L^2$. Thus, strictly speaking, we are generating discrete
spatiotemporally colored random fields that are derived from the
model~\eqref{eq:coloredspattemp} without being approximate solutions
of the continuous problem.

The spatiotemporal random field~$\zeta(\vec{r},t)$ is simulated by transforming a
discretised version~$\zeta_{ij}(t)$ of~\eqref{eq:coloredspattemp} to
Fourier space:
\begin{equation}
  \label{eq:zetafourier}
  \zeta_{\mu\nu}(t) = (\Delta x)^2\sum_{i,j = 1}^L \exp (-\mathbf{i} k\cdot \vec{r}) \zeta_{ij} (t)
\end{equation}

We denote the discrete Fourier transform~$\zeta_{\mu\nu}(t)$ with greek indices~$\mu$, $\nu$ rather than~$i$, $j$ and 
the coordinate~$k$ in frequency space is

\[
k=\frac{2 \pi}{L \Delta x} (\mu, \nu), \qquad \mu, \nu=0, \dots, L-1.
\]

Then the Fourier transformed field~$\zeta_{\mu\nu}(t)$ at the next
time step $t + \Delta t$ is calculated by
\begin{align}
  \label{eq:spattempnext}
  \zeta_{\mu \nu} (t+\Delta t) &=   \zeta_{\mu \nu} (t) \exp \left(- \frac{c_{\mu\nu}}{\tau} \Delta t\right) + \sqrt{\frac{\epsilon (L \Delta x)^2}{\tau c_{\mu \nu}} \left[ 1- \exp \left( - \frac{c_{\mu\nu}}{\tau} \Delta t \right) \right]} \alpha_{\mu \nu}
\end{align}

Here, $\alpha_{\mu \nu}$ is the Fourier transform of an uncorrelated
Gaussian noise field~$\alpha_{i j}$. The efficiency of this method is
increased by directly generating the Fourier transformed
field~$\alpha_{\mu \nu}$. The (complex-valued) discrete Fourier
transform~$\alpha_{\mu \nu}$ of real-valued fields~$\alpha_{ij}$ obeys
some symmetries that lead to the following restrictions:
\begin{align}
  \label{eq:alphasym}
\alpha_{\mu \nu} & = \alpha_{L - \mu, L - \nu}^*\\
\alpha_{\mu\nu} & \in \mathbb{R}, \quad \text{for } \mu, \nu= 0, \frac{L}{2}
\end{align}

where $z^*$ denotes the complex conjugate of~$z \in \mathbb{C}$. The
condition\,(\ref{eq:alphasym}) means that~$\alpha_{\mu \nu}$ have the
same real part as~$\alpha_{L -\mu, L-\nu}$ found by reflecting through
the centre ($\mu=\nu=1/2$) whereas the imaginary parts only differ by
opposite signs. 

Also in~\eqref{eq:spattempnext}, $c_{\mu \nu}$ is the Fourier transform of the discretisation of the differential 
operator ~$\mathcal{L} = \id - \lambda^2 \Delta$:
\begin{equation}
  \label{eq:cTrans}
  c_{\mu \nu} = 1 - \frac{2 \lambda^2}{(\Delta x)^2} \left[ \cos \left( \frac{2\pi \mu}{L} \right) + \cos \left( \frac{2\pi \nu}{L} \right) - 2 \right].
\end{equation}

For further details on the derivation of this method, we refer the reader to \cite{Gar:99} or \cite{Gar:92a}.

\setcounter{equation}{0}
\section{Resident-invader competition-diffusion model with infected invader in a variable environment}
\label{sec:results}

\subsection{Local dynamics with multiplicative white noise and induced transitions}
\label{sec:locnoise}

\begin{figure}[ht]
\centering
\includegraphics[width=0.32\textwidth]{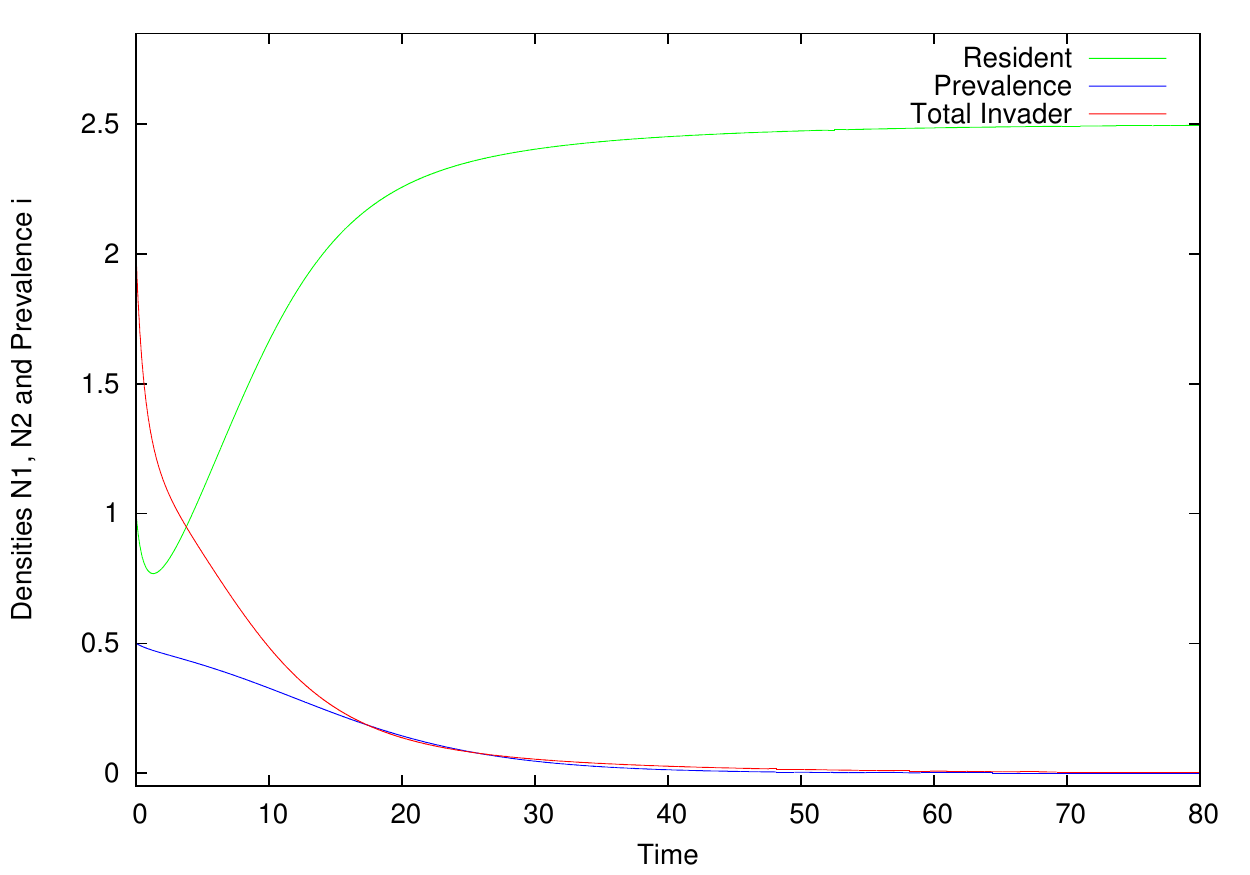}
\includegraphics[width=0.32\textwidth]{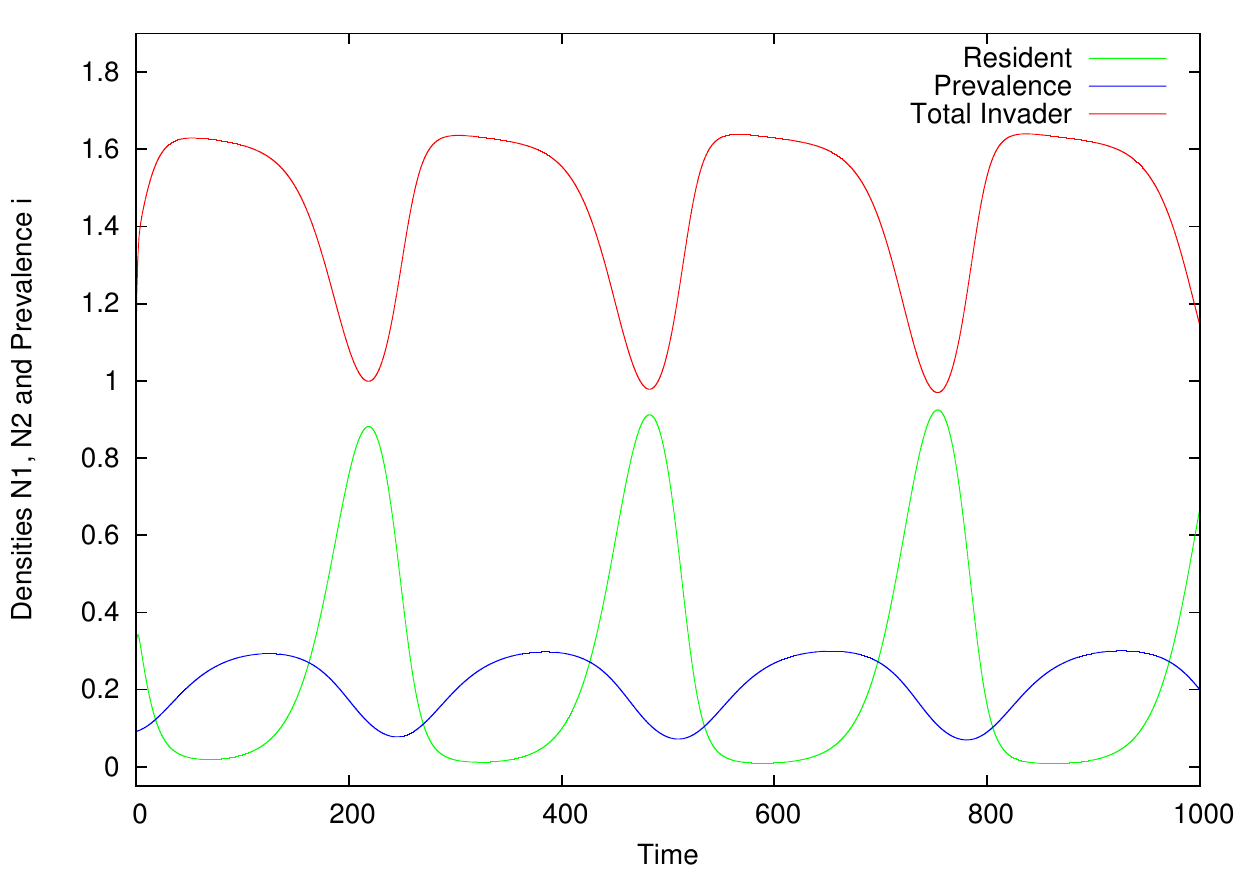}
\caption{Bistability of resident-only state and oscillating coexistence}
\label{fig:homogbistab}
\end{figure}

Not surprising and like for $q=0$ \cite{Ven:01a, Dri:04}, Lehmann \cite{Leh:11} and Woyzichovski \cite{Woy:13} found disease-induced oscillations for $q=1$ as well. Their interesting result was that there may be bistability of the resident-only state and the limit cycle when coexisting resident, susceptible and infected invaders oscillate, cf. Figure\,\ref{fig:homogbistab}. The following parameters have been used for the latter setting:
\begin{align}
\label{eq:parabinoise}
                      r_1&=1.2500\,,~r_2=1.0000\,,~r_3=0.6775\nonumber\\
 c_{11}=0.5000\,, ~c_{12}&=0.8500\,,~c_{21}=0.4250\,,~c_{13}=0.5000\,,~c_{31}=0.5000\\
                   c_{22}&=0.6000\,,~c_{23}=0.4000\,,~c_{32}=0.6010\,,~c_{33}=0.5000\nonumber\\
                  \lambda&=0.4000\,,~\mu=0.0250\nonumber
\end{align}

All other semi-trivial states turn out to be unstable for this parameter range.

Now, one effect of white noise in locally multiple stable systems is shown, i.e., the switch from one stable attractor to the other for sufficiently but not too high noise intensity. For simplicity, the linear density dependence (\ref{eq:linearnoise}) of the intensity is chosen, which has been successfully applied to numerous cases. Here, the leaving of the initial limit cycle is demonstrated.\\

\begin{figure}[ht]
\centering
\includegraphics[width=0.32\textwidth]{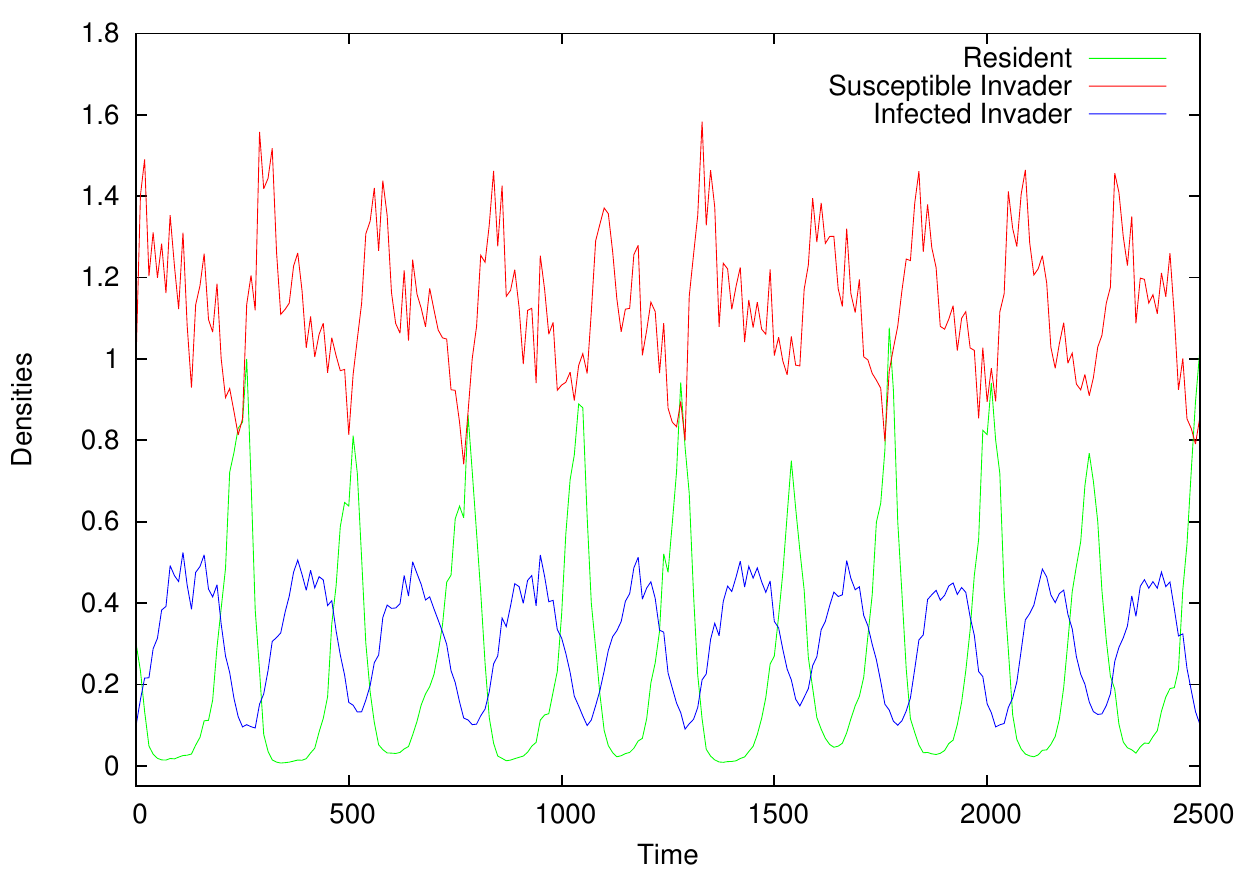}
\includegraphics[width=0.32\textwidth]{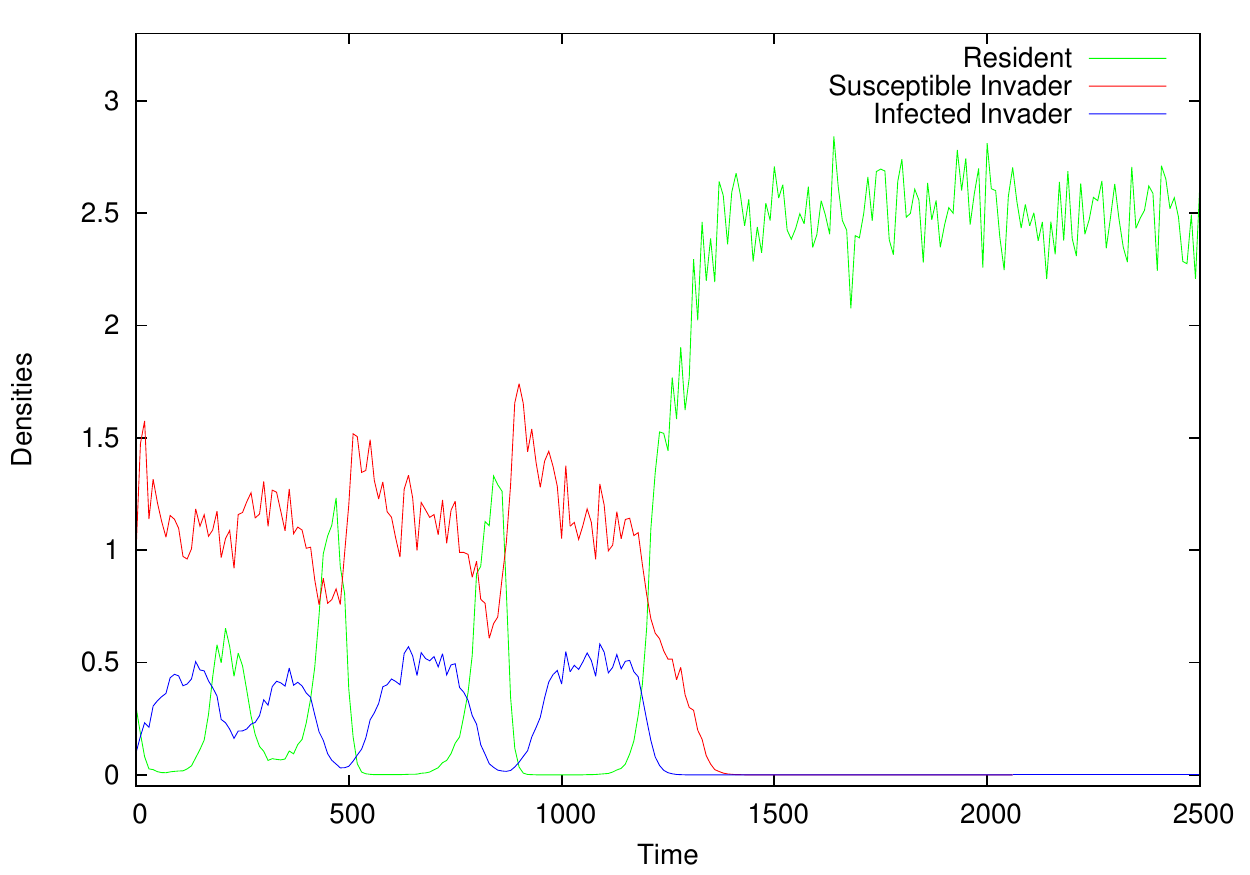}
\includegraphics[width=0.32\textwidth]{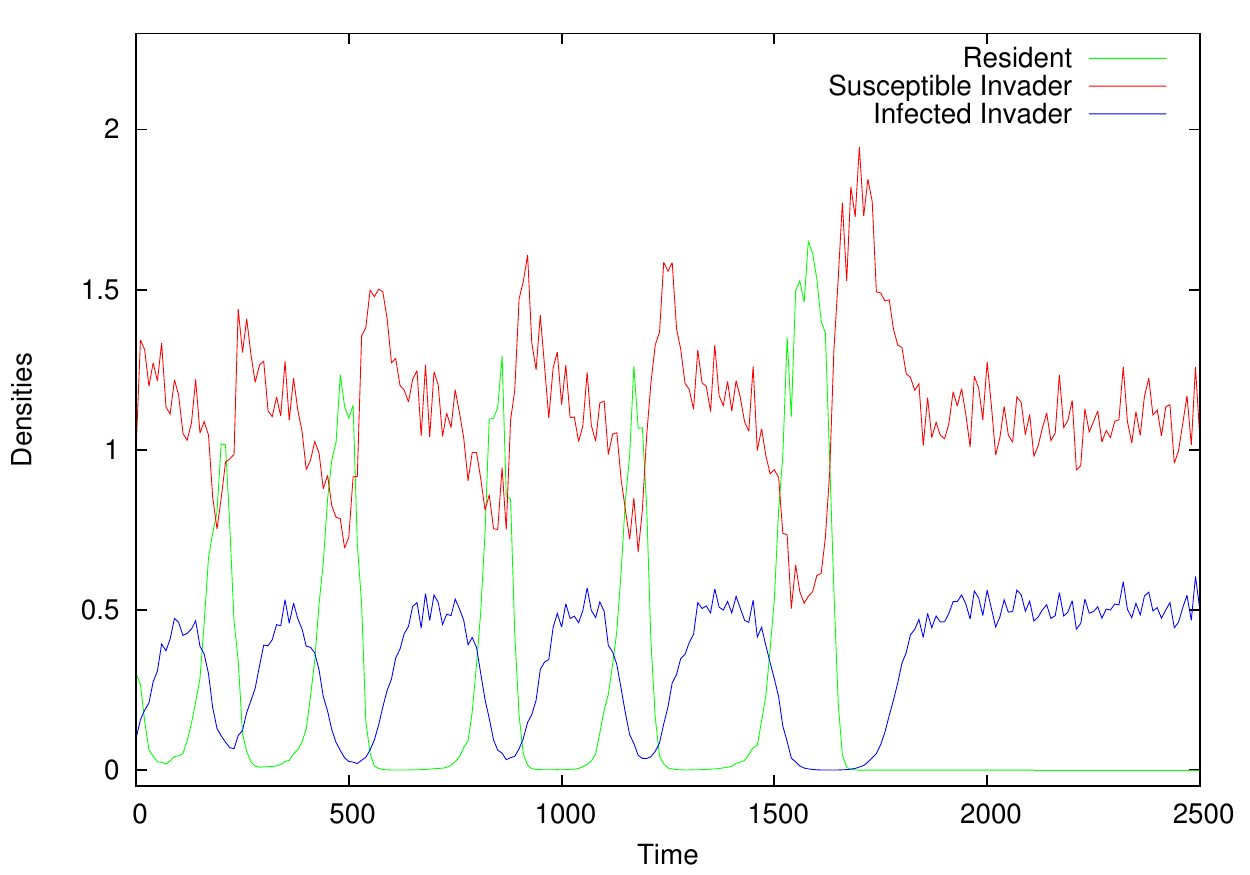}
\caption{Noise-induced transitions from oscillating coexistence to resident-only state resp. resident extinction. Linear density dependence (\ref{eq:linearnoise}) with $\omega_{ii}=0.1\, ~i=1,2,3$\,.}
\label{fig:switchbistab}
\end{figure}

The three subfigures of Figure\,\ref{fig:switchbistab} show typical outcomes of hundreds of simulations with different seeds of the random number generator. The left subfigure shows the persistence of the limit cycle whereas the middle demonstrates the expected leaving of the cycle for the other stable stationary solution, i.e., the resident-only state. The result in the right subfigure appeared a bit unexpected, however, due to some catastrophic shift, the resident died out and the remaining susceptible and infected invader survived. One should have in mind that the latter $(0, X_2^s, X_3^s)$ state is unstable to the reintroduction of the resident.

\subsection{Invasions and noise I}
\label{sec:ccsi}

\subsubsection{Linear noise and biocontrol of invasion}
\label{sec:whitets1a}

For the beginning, the results in the mentioned previous papers \cite{Mal:11,Mal:13}, where simulated landslides led to bare land competitively re-invadable by resident and invader, are reproduced with external noise. The initial condition is a ``red'' invader patch at its emergent carrying capacity $r_{22}/c_{22}$ at the ``upper left corner'' of the ``green'' habitat of the native species at $r_{11}/c_{11}$. This patch should exceed the related critical patch size. Otherwise it will simply decay regardless of its competitive strength and mobility \cite{Nit:74, Mal:85a}. Zero-flux boundary conditions are applied.\\

The linear noise (\ref{eq:linearnoise}) and parameters from the previous publications have been taken:
\begin{align}
\label{eq:paraold}
                      r_1&=1.000\,,~r_2=1.000\,,~r_3=0.800\nonumber\\
  c_{11}=1.000\,, ~c_{12}&=1.300\,,~c_{21}=1.200\,,~c_{13}=1.299\,,~c_{31}=1.201\nonumber\\
                   c_{22}&=0.999\,,~c_{23}=0.998\,,~c_{32}=1.001\,,~c_{33}=1.000\\
                  \lambda&=0.405\,,~\mu=0.200\nonumber\\
              \omega_{ii}&=0.250\,; ~D_i=45.000\,; ~i=1,2,3.\nonumber
\end{align}

The invader patch spreads and seems to grow unstoppable.
\begin{figure}[!ht]
\center
\subfigure[t=240]{\includegraphics[width=0.19\textwidth]{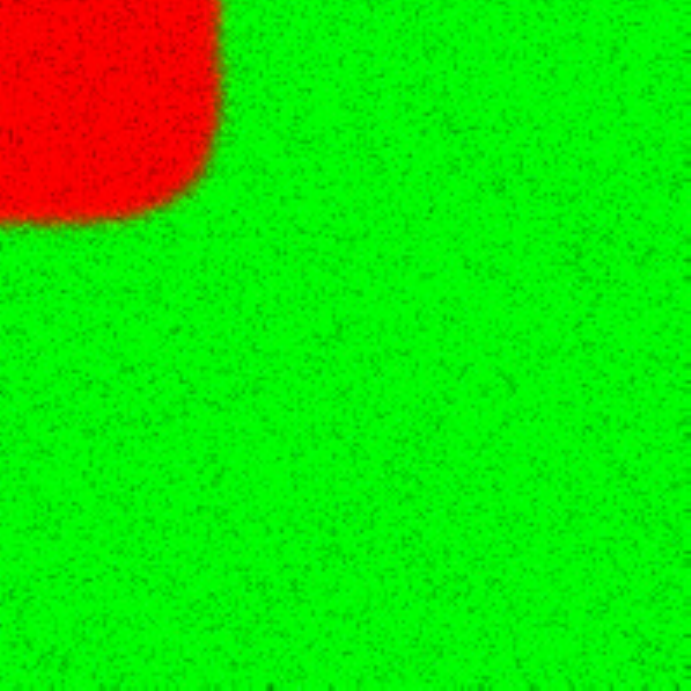}}
\subfigure[480]{\includegraphics[width=0.19\textwidth]{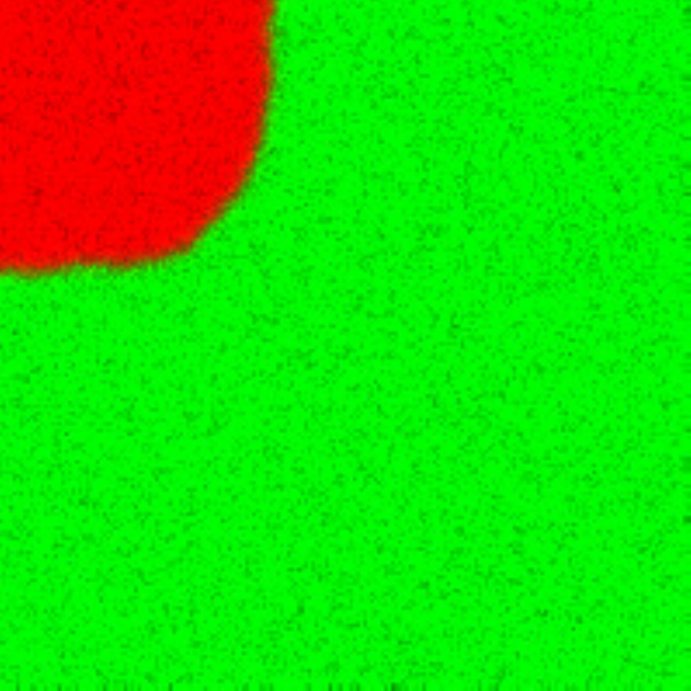}}
\subfigure[720]{\includegraphics[width=0.19\textwidth]{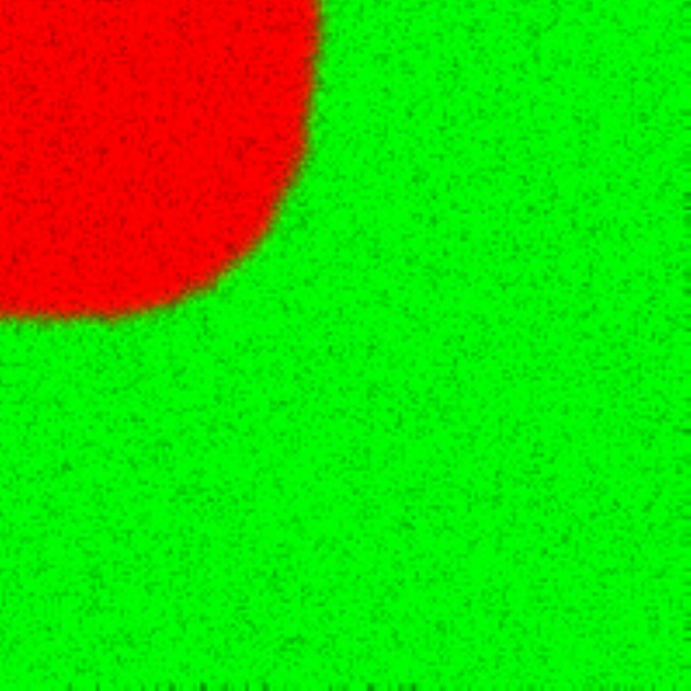}}
\subfigure[960]{\includegraphics[width=0.19\textwidth]{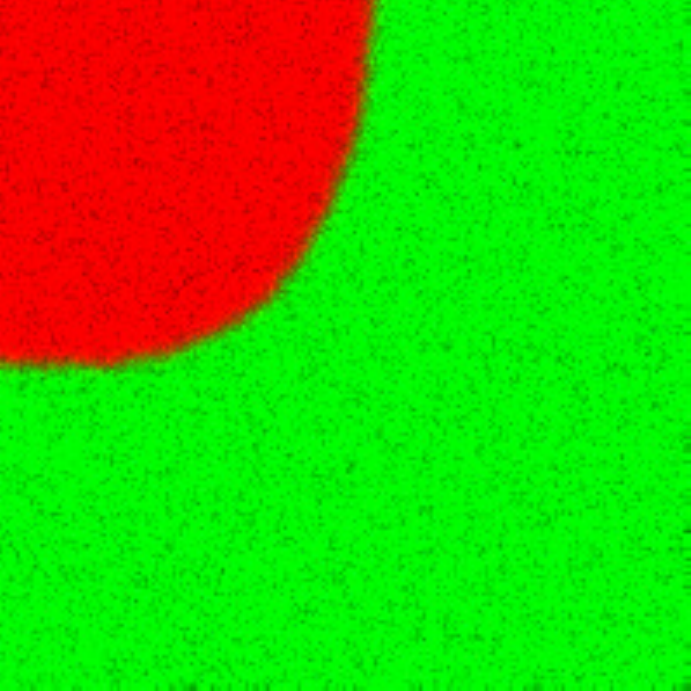}}
\subfigure[1190]{\includegraphics[width=0.19\textwidth]{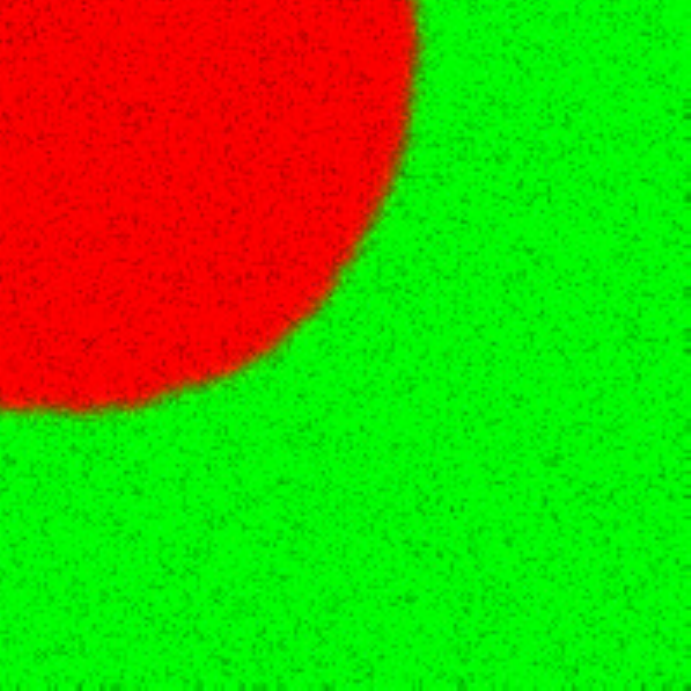}}\\
\subfigure[t=240]{\includegraphics[width=0.19\textwidth]{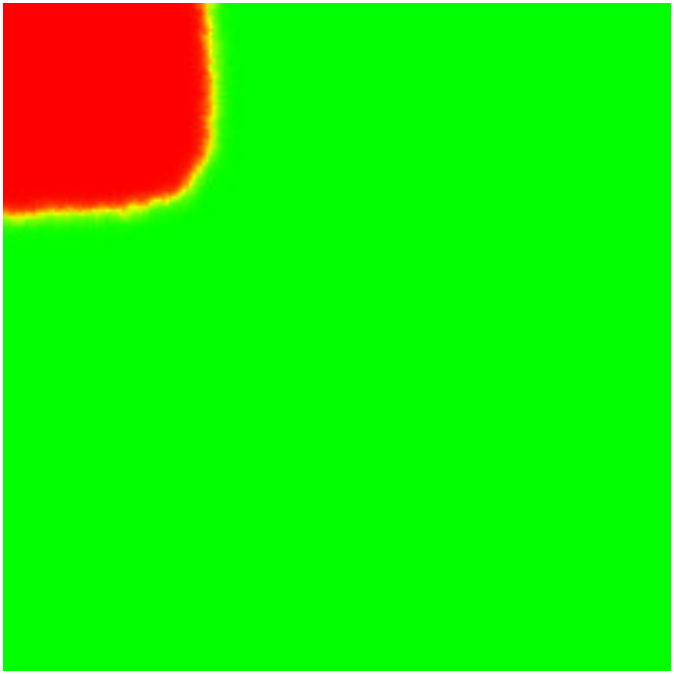}}
\subfigure[480]{\includegraphics[width=0.19\textwidth]{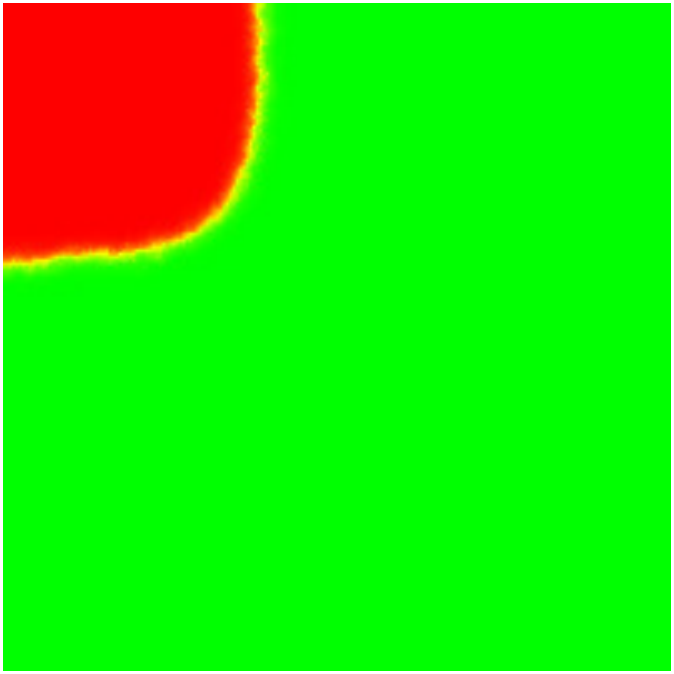}}
\subfigure[720]{\includegraphics[width=0.19\textwidth]{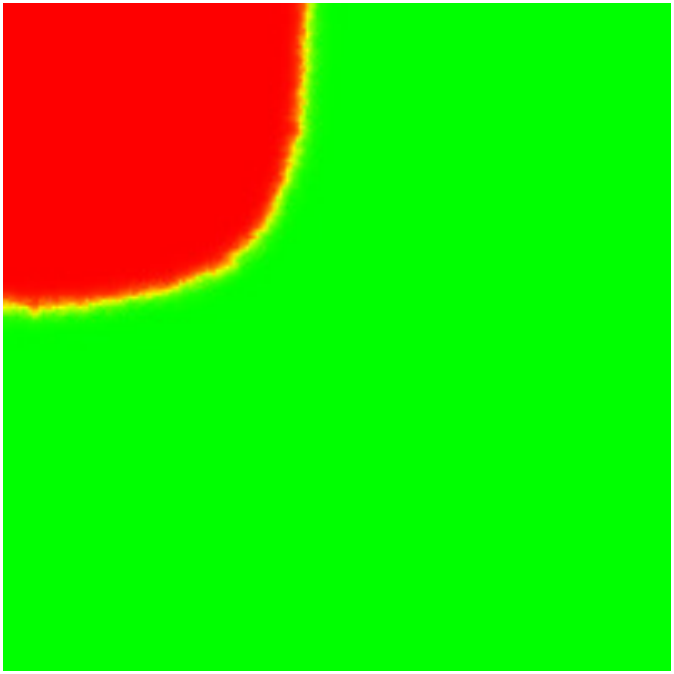}}
\subfigure[960]{\includegraphics[width=0.19\textwidth]{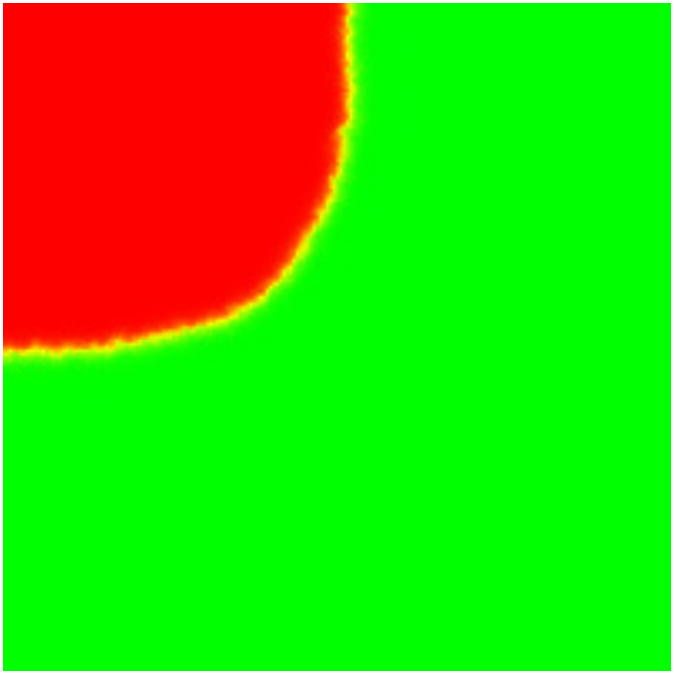}}
\subfigure[1190]{\includegraphics[width=0.19\textwidth]{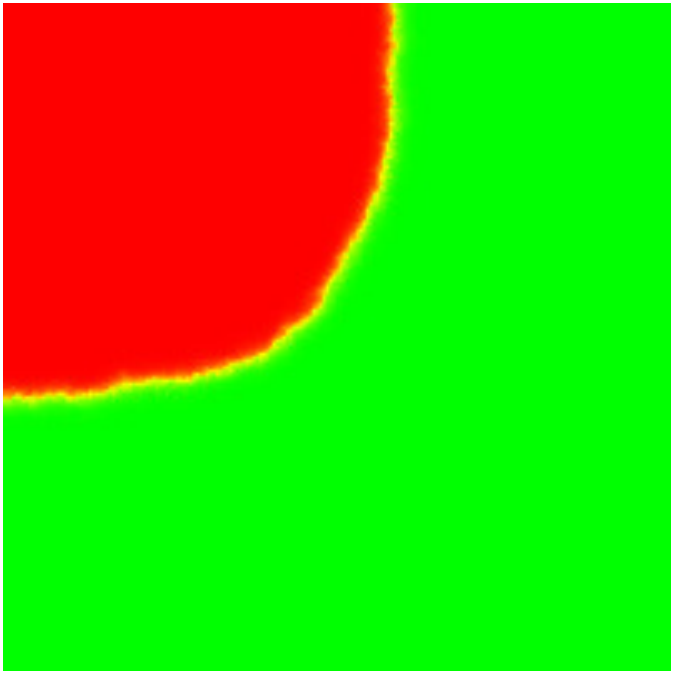}}
\caption{Simulation 1: Parameters as given in (\ref{eq:paraold}). The first row shows results with white 
noise, the second row shows spatiotemporally colored noise with~$\omega_{ii}=0.03$, $\epsilon=0.001$ and 
correlation lengths~$\tau=1$ and~$\lambda=1$.}
\label{fig:simpar01}
\end{figure}

Then, a biological control measure is applied. The invader population is partly infected and the invasion successfully rolled back.
\begin{figure}[!ht]
\center
\subfigure[t=1200]{\includegraphics[width=0.19\textwidth]{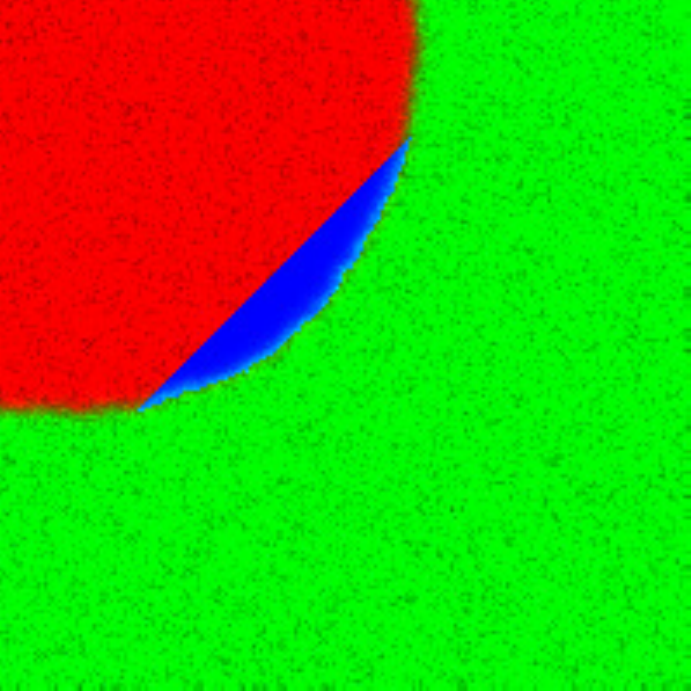}}
\subfigure[1400]{\includegraphics[width=0.19\textwidth]{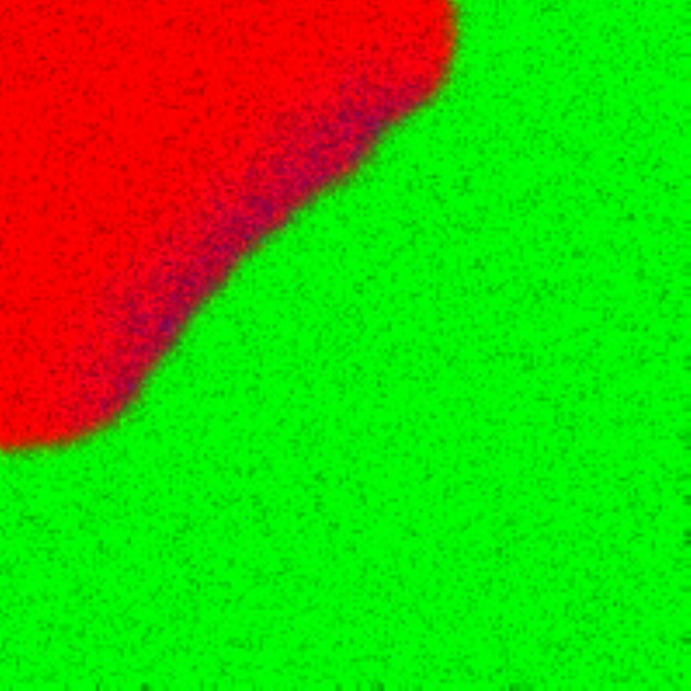}}
\subfigure[1600]{\includegraphics[width=0.19\textwidth]{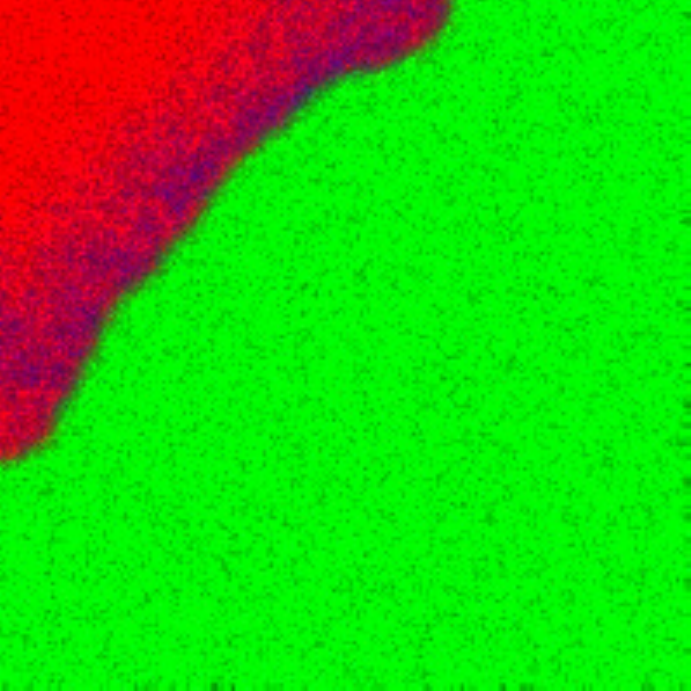}}
\subfigure[1800]{\includegraphics[width=0.19\textwidth]{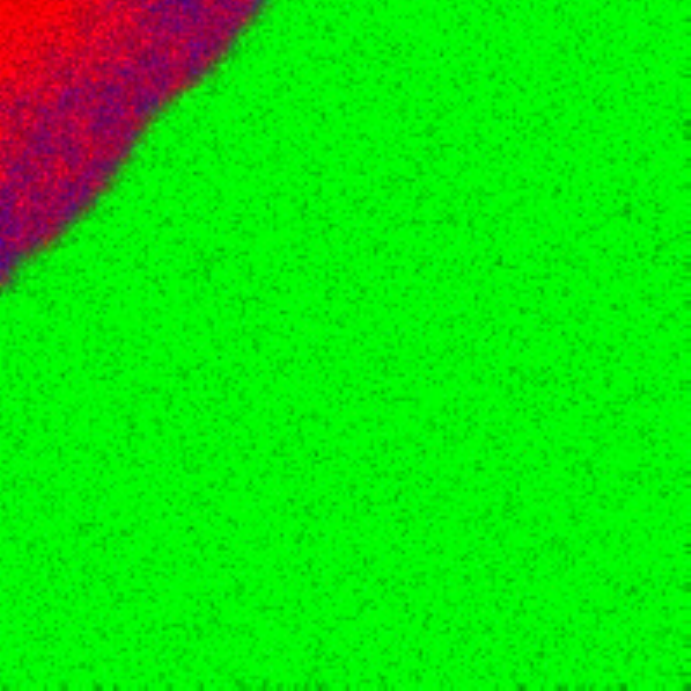}}
\subfigure[2000]{\includegraphics[width=0.19\textwidth]{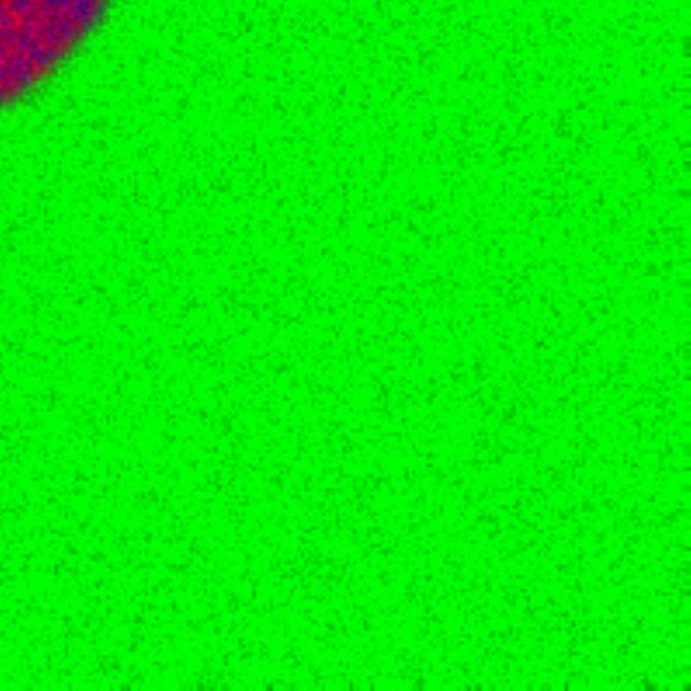}}
\caption{Simulation 1 continued: Partial ``blue'' infection of the invader at t=1200.}
\label{fig:simpar02}
\end{figure}

\begin{figure}[!ht]
\center
\includegraphics[width=0.75\textwidth]{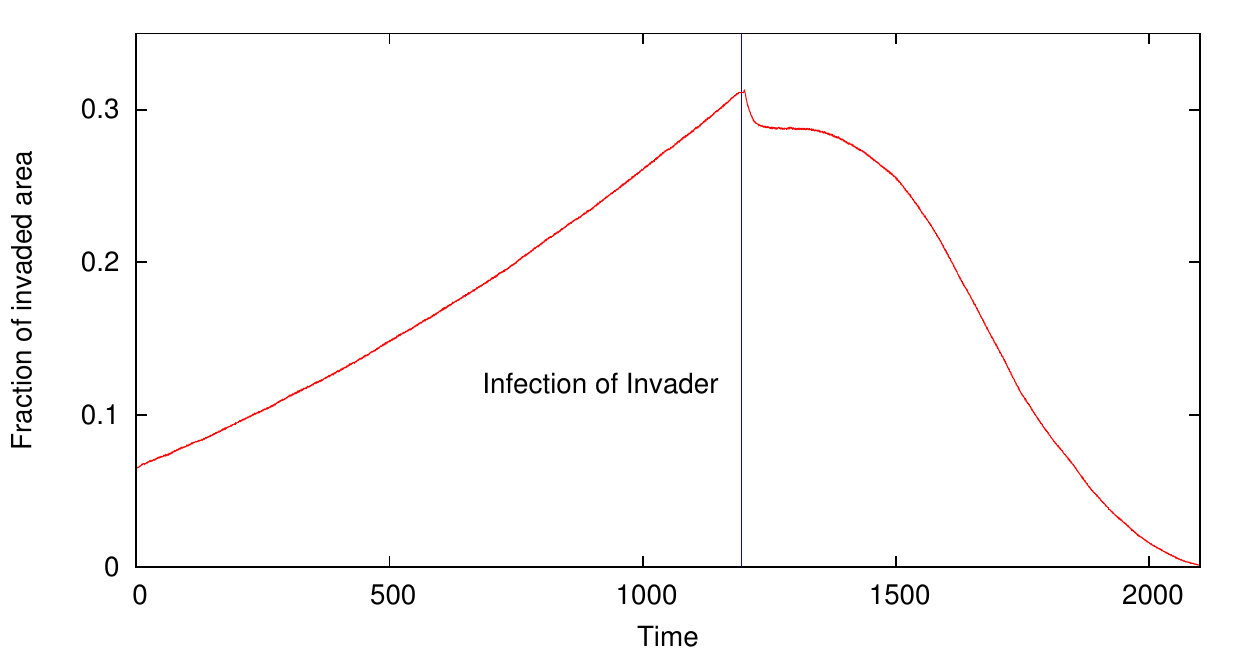}
\caption{Fraction of invaded area before and after partial infection.}
\label{fig:invareainfect}
\end{figure}

The changes of the fraction of the invaded area can be seen in Figure\,\ref{fig:invareainfect}.

\subsubsection{Nonlinear response to noise and noise control of invasion}
\label{sec:whitets1b}

We now consider the saturating response to noise described above (\ref{eq:simpnoise}) with
\begin{align}
\label{eq:paranonlinearnoise}
 m=n=2\,; ~\omega_{11}&=50\,, ~\omega_{22}=0.1\,, ~\omega_{33}=10\,;\nonumber\\
          ~\gamma_{11}&=100\,, ~\gamma_{22}=3\,, ~\gamma_{33}=300.
\end{align}

\begin{figure}[!ht]
\center
\subfigure[t=100]{\includegraphics[width=0.19\textwidth]{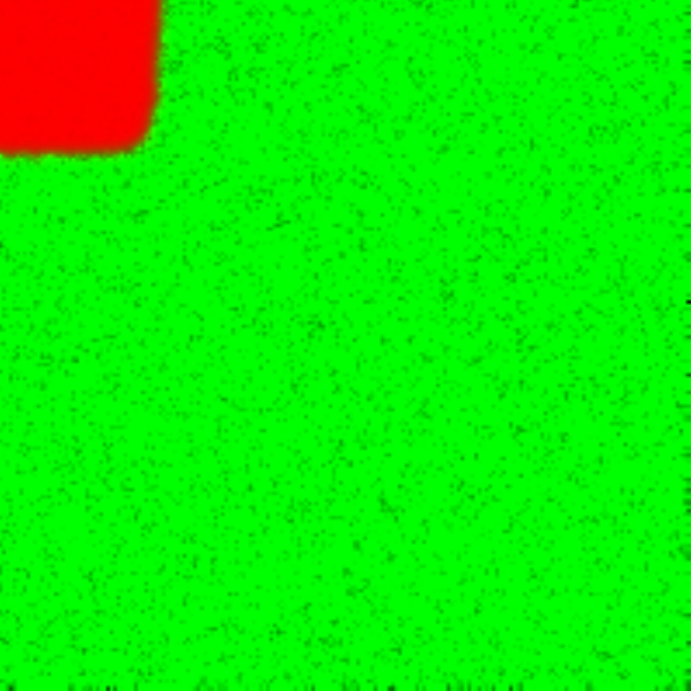}}
\subfigure[300]{\includegraphics[width=0.19\textwidth]{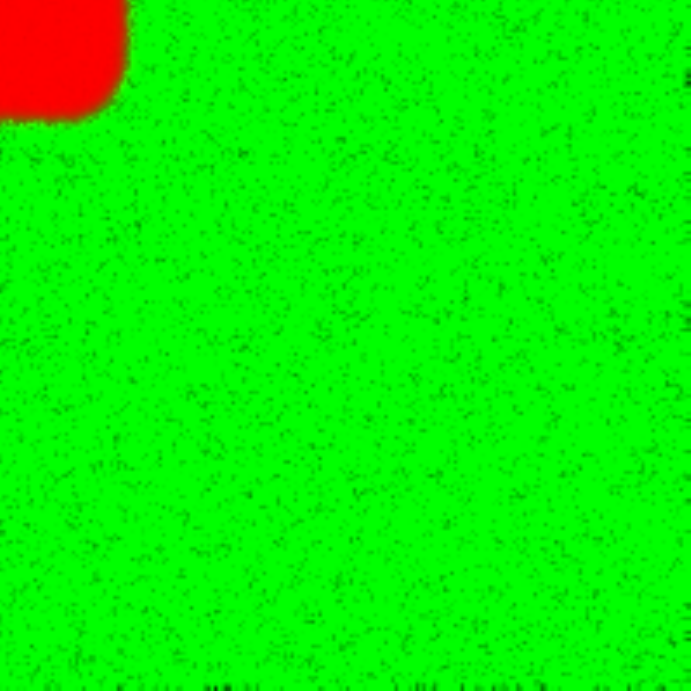}}
\subfigure[400]{\includegraphics[width=0.19\textwidth]{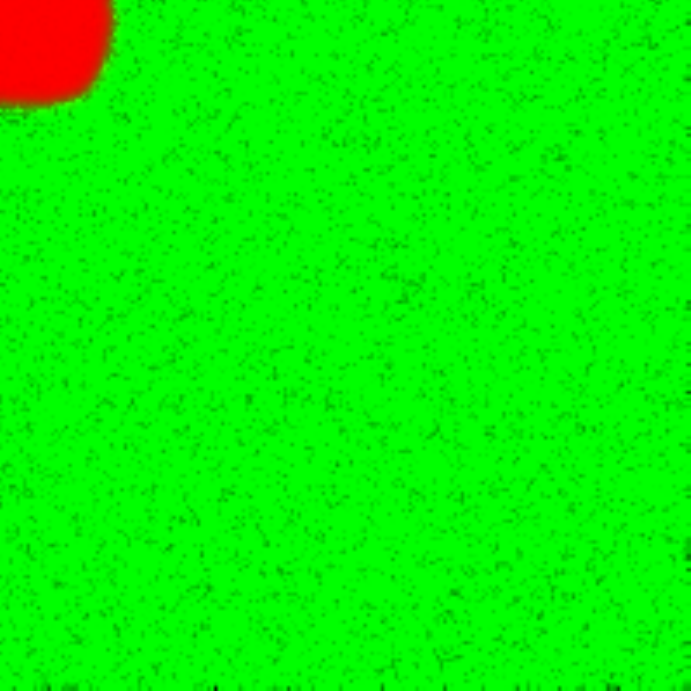}}
\subfigure[500]{\includegraphics[width=0.19\textwidth]{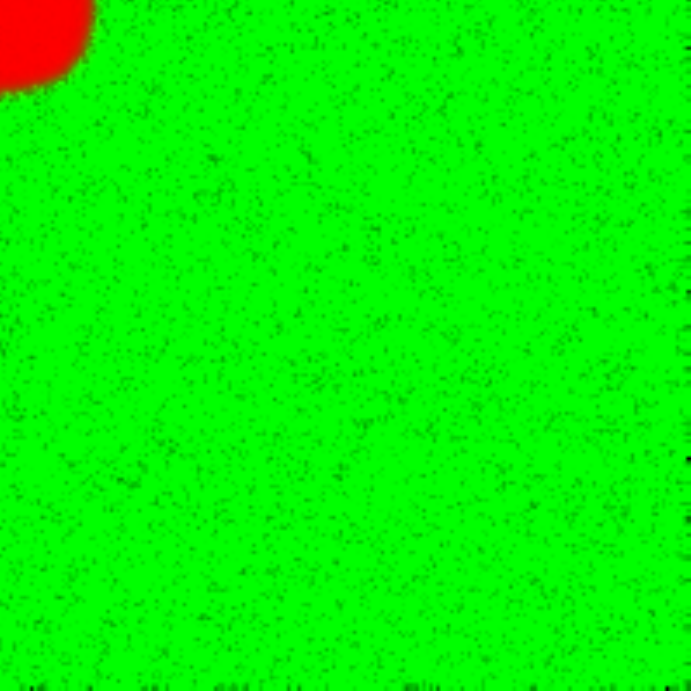}}
\subfigure[700]{\includegraphics[width=0.19\textwidth]{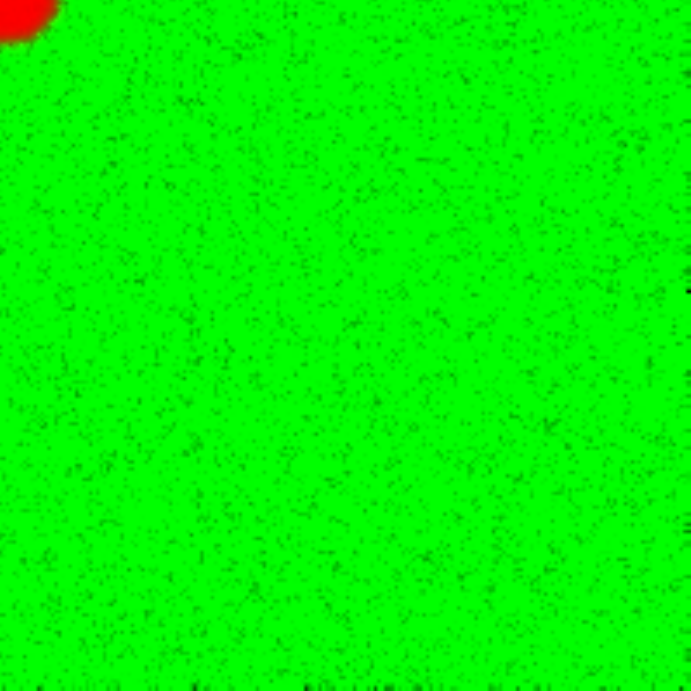}}\\
\subfigure[t=500]{\includegraphics[width=0.19\textwidth]{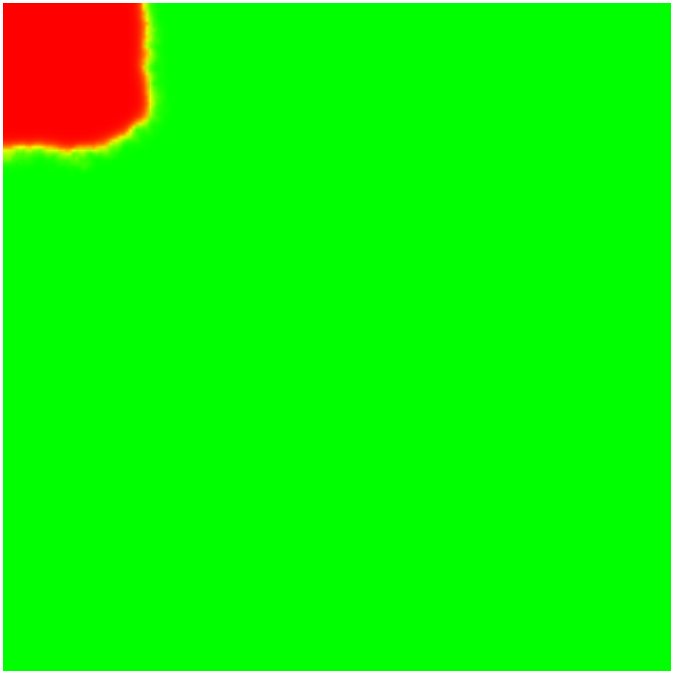}}
\subfigure[1000]{\includegraphics[width=0.19\textwidth]{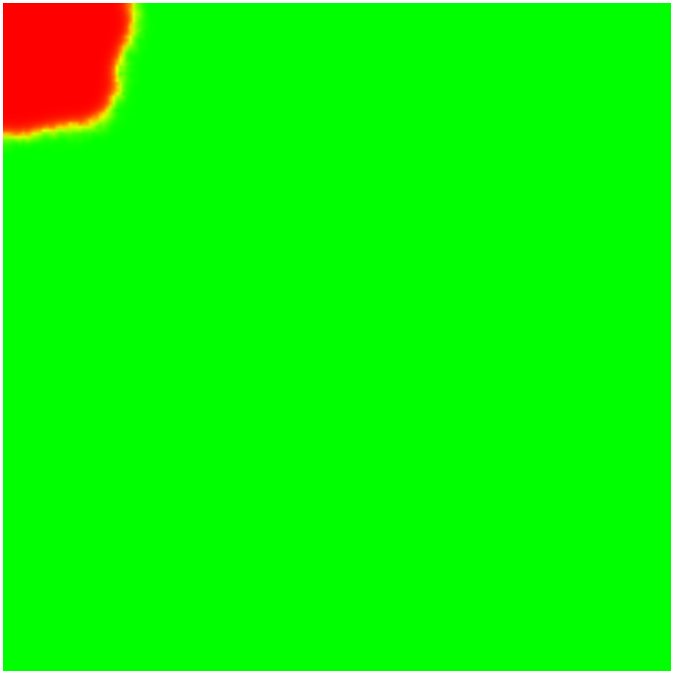}}
\subfigure[1500]{\includegraphics[width=0.19\textwidth]{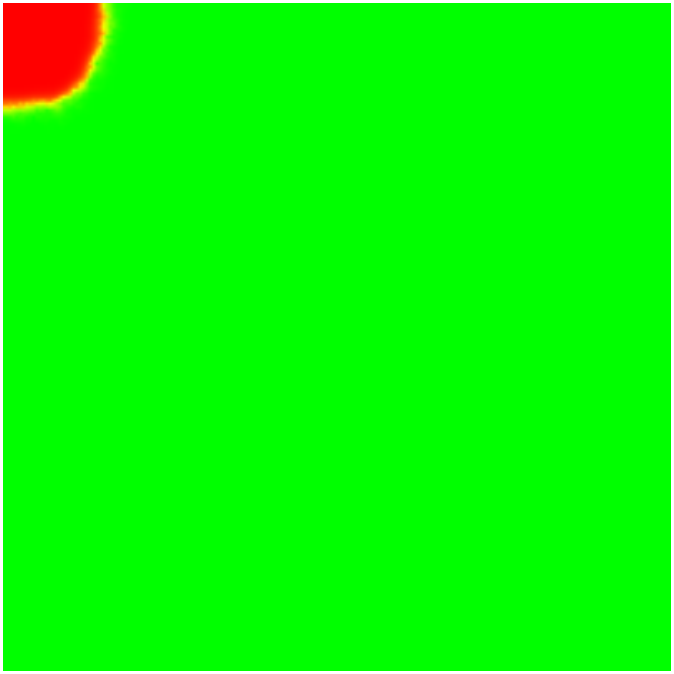}}
\subfigure[2000]{\includegraphics[width=0.19\textwidth]{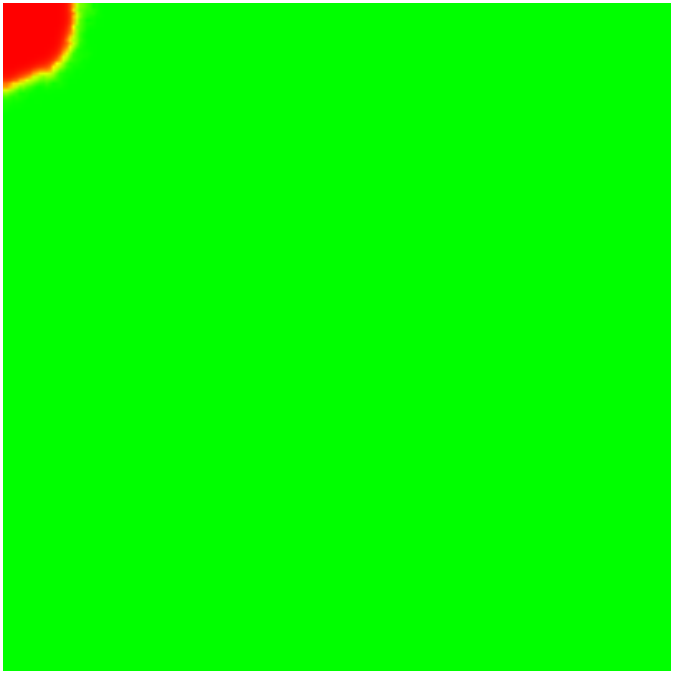}}
\subfigure[2500]{\includegraphics[width=0.19\textwidth]
{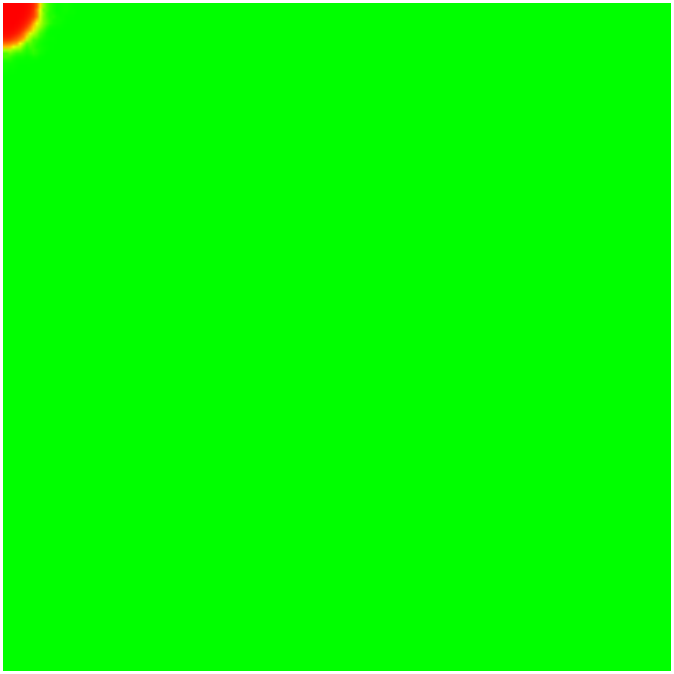}}
\caption{Repressed invasion through population-specific nonlinearly density-dependent noise intensities for white noise
in the first row. In the second row for colored noise with correlation lengths~$\tau=1$ and $\lambda=1$, $\omega_{11}=0.05$,
$\omega_{22}= \omega_{33}=0.01$ qualitatively similar behaviour is observed, albeit on a much slower timescale. 
However, other simulations have shown that stronger temporal and spatial correlations can support invasions.}
\label{fig:simpar03}
\end{figure}

The extinction of the invader due to hostile environmental conditions is shown in Figure~\ref{fig:simpar03}. The 
resident is used and adapted to the environment and happily survives.\\

This dynamics is only due to the specific noise response of the populations. All growth and interaction parameters remained the same as in 
sec.\,\ref{sec:whitets1a}

\subsection{Invasions and noise IIa}
\label{sec:whitets2}

Coming back to the parameter range of bistability of resident-only state and oscillating coexistence of all three populations, i.e., parameters (\ref{eq:parabinoise}), the initial condition is chosen as uniformly populated by the resident $X_1$ at its emergent carrying capacity. At a defined location at the boundary, an initial patch of the invading populations $X_2$ and $X_3$ attempts to spread. Again, simply linear noise (\ref{eq:linearnoise}) is applied and the influence of increasing noise intensity $\omega_{ii}=\omega\,; ~i=1,2,3$\,; studied.\\

\begin{figure}[!ht]
\center
\includegraphics[width=0.75\textwidth]{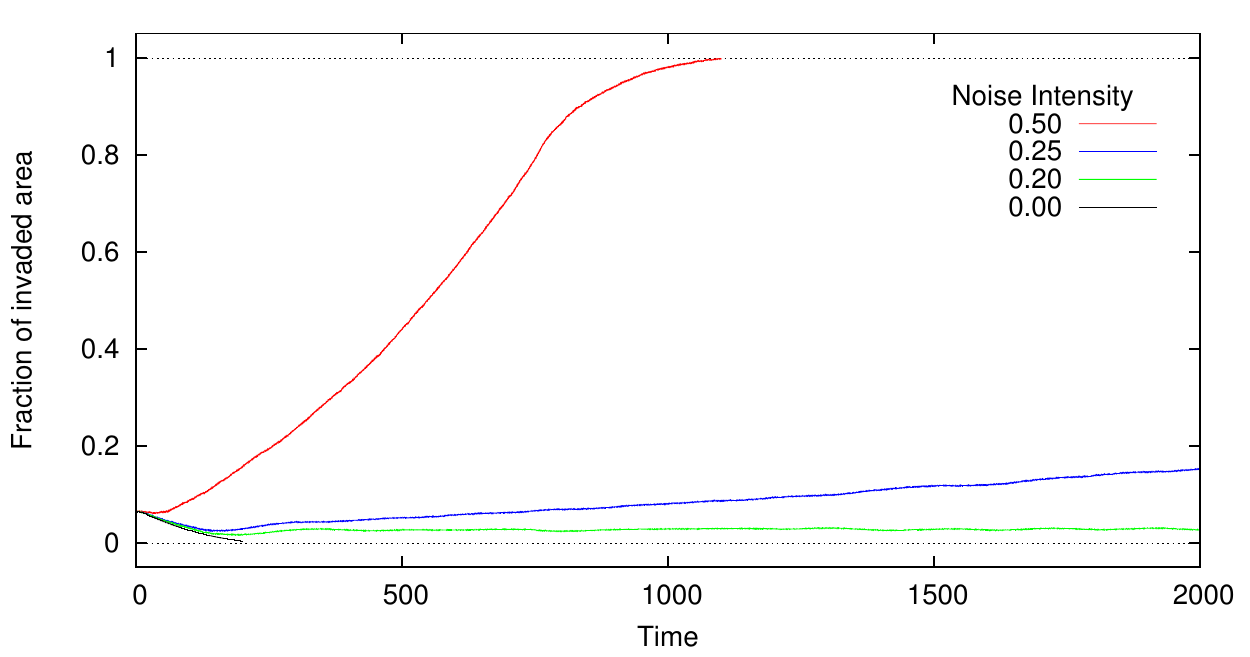}
\caption{Decline resp. growth of invaded area for increasing white noise intensity.}
\label{fig:invareaoscilla1}
\end{figure}

It is seen in Figure\,\ref{fig:invareaoscilla1} that a successful invasion requires a certain supercritical noise intensity.

\begin{figure}[!ht]
\center
\subfigure{\includegraphics[width=0.19\textwidth]{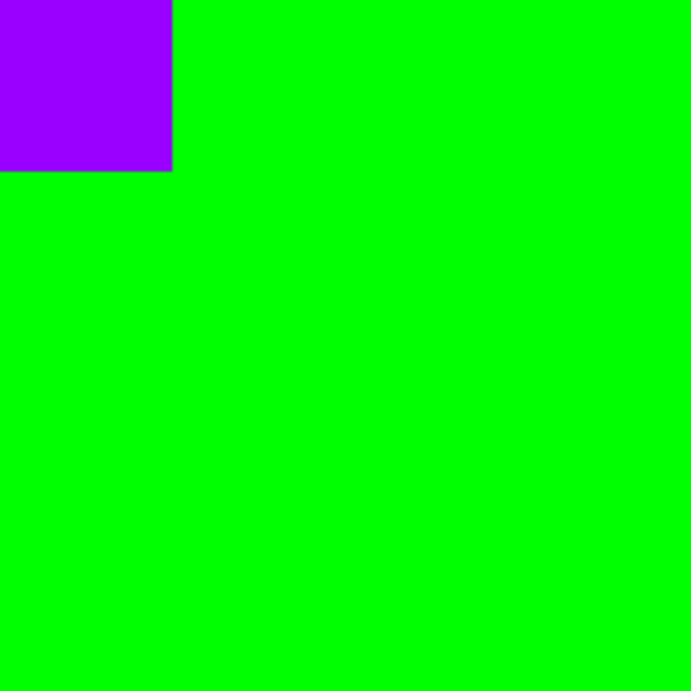}}
\subfigure{\includegraphics[width=0.19\textwidth]{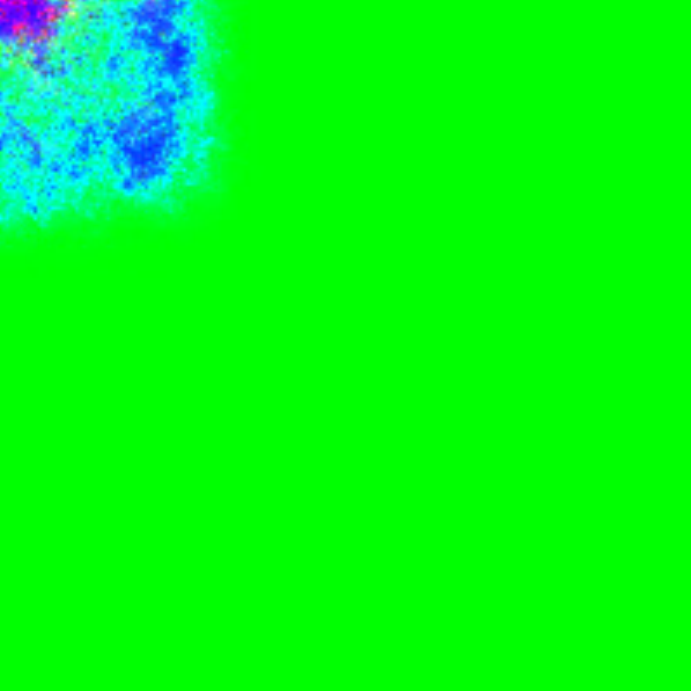}}
\subfigure{\includegraphics[width=0.19\textwidth]{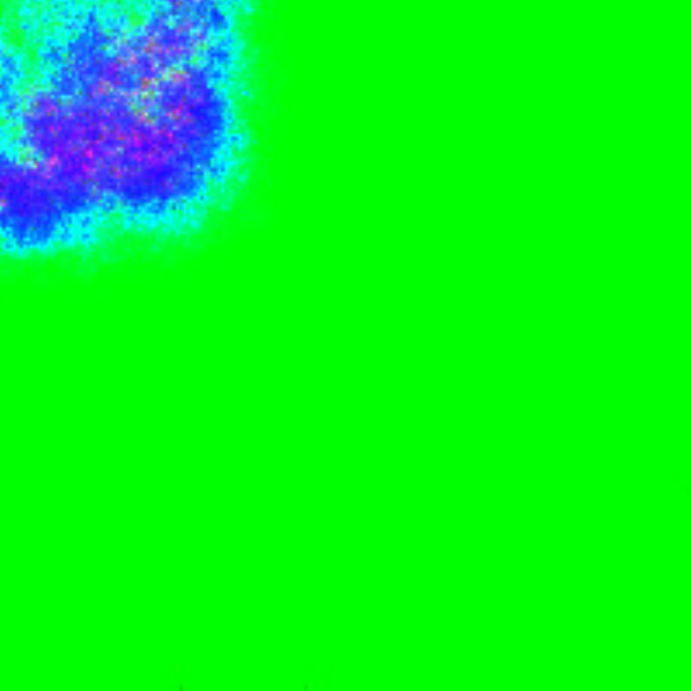}}
\subfigure{\includegraphics[width=0.19\textwidth]{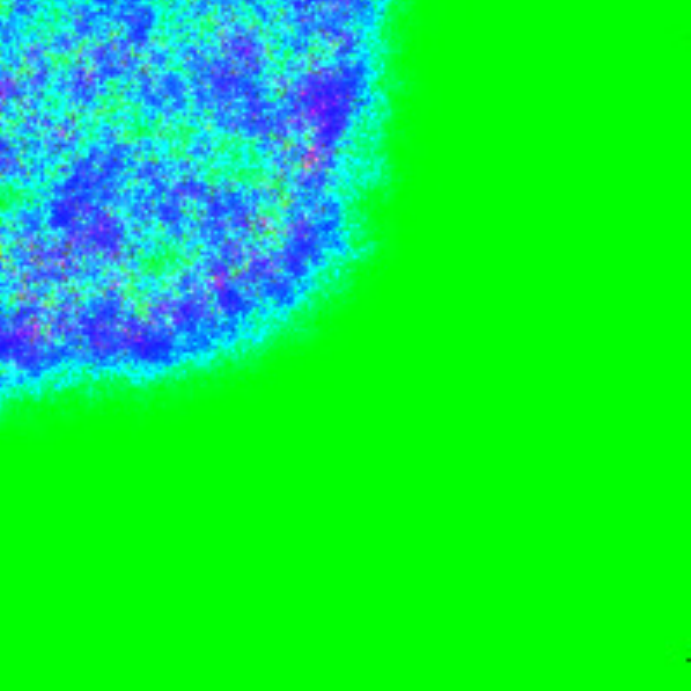}}
\subfigure{\includegraphics[width=0.19\textwidth]{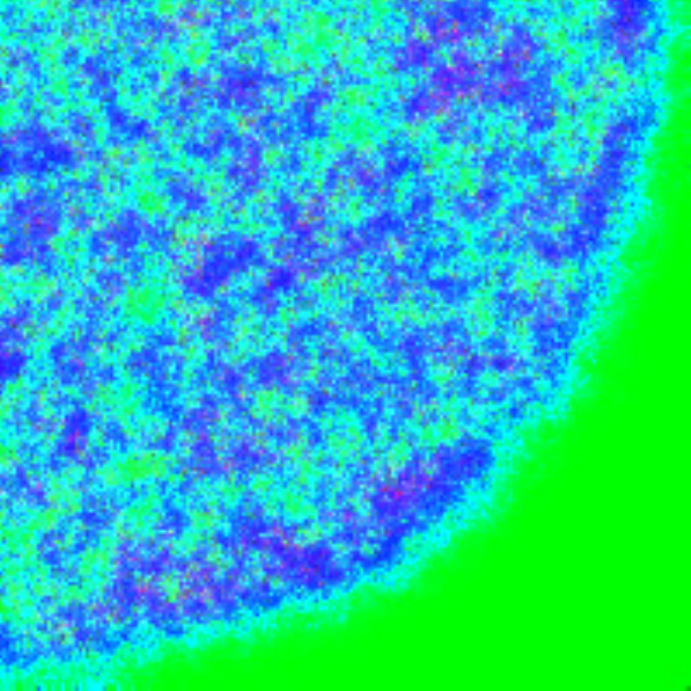}}\\
\subfigure[t=0]{\includegraphics[width=0.19\textwidth]{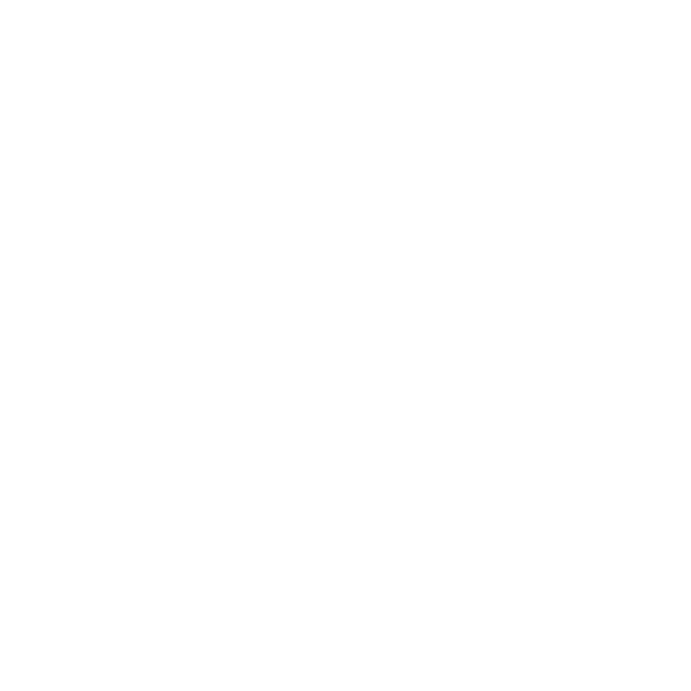}}
\subfigure[100]{\includegraphics[width=0.19\textwidth]{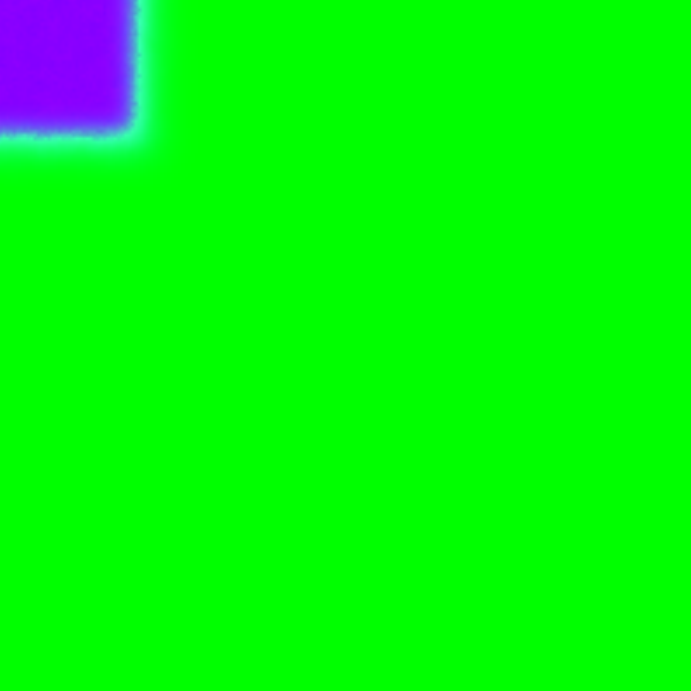}}
\subfigure[150]{\includegraphics[width=0.19\textwidth]{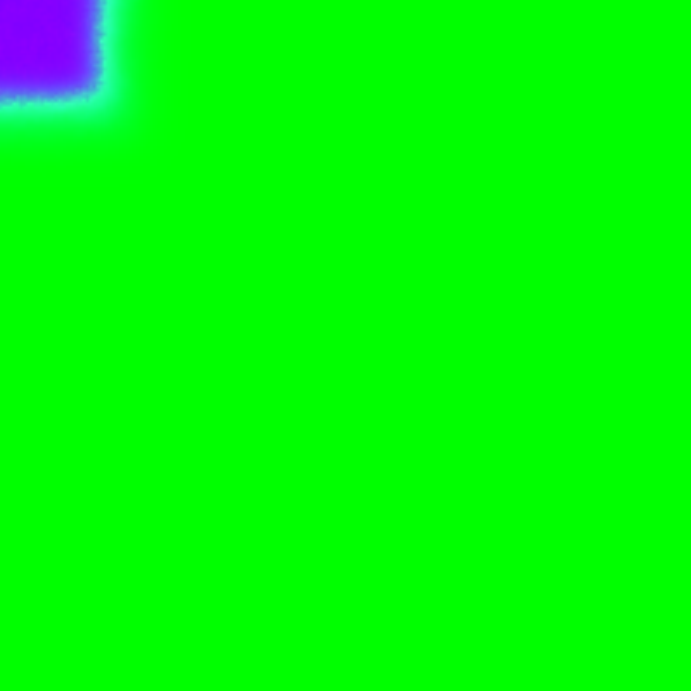}}
\subfigure[350]{\includegraphics[width=0.19\textwidth]{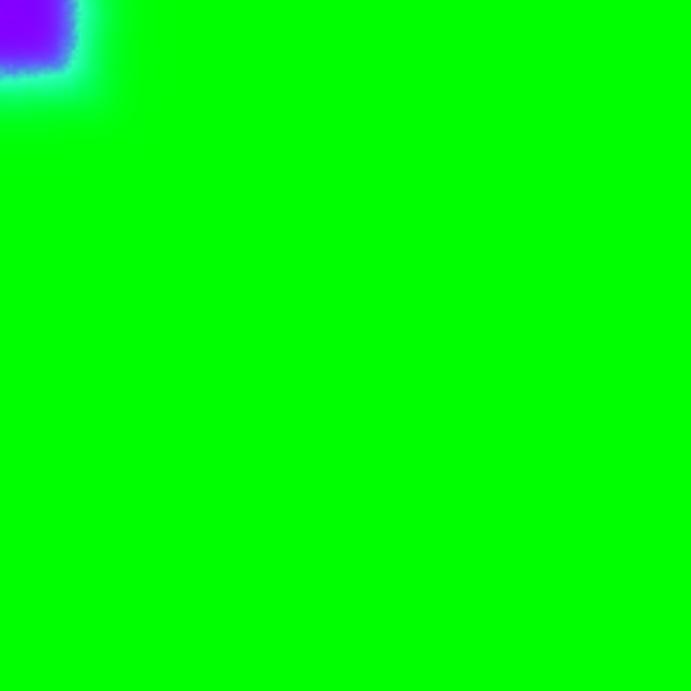}}
\subfigure[750]{\includegraphics[width=0.19\textwidth]{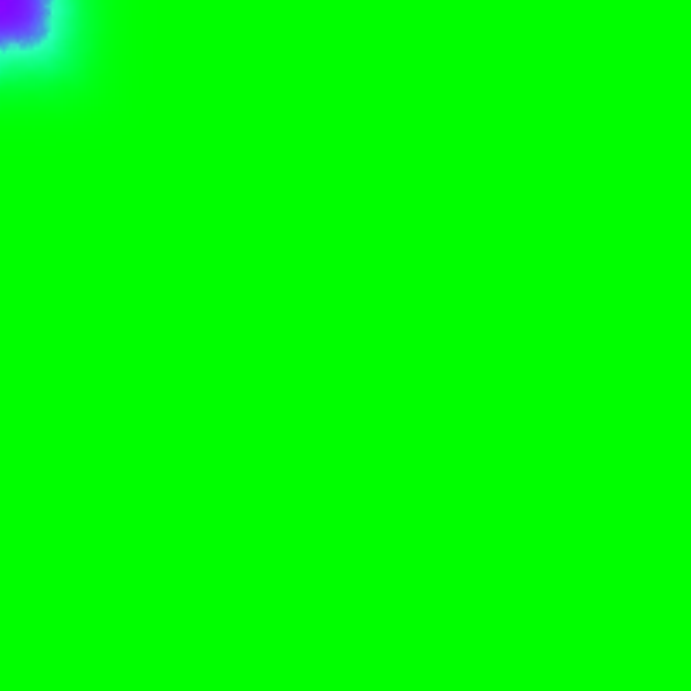}}
\caption{Upper row: Successful fast invasion of the resident's area for linear white noise (\ref{eq:linearnoise}) with $\omega_{ii}=0.5\,; 
~i=1,2,3$\,. Lower row: Stopped invasion of the resident's area for nonlinear noise (\ref{eq:paranonlinearnoise}) with 
$\omega_{ii}=0.25\,;~i=1,2,3\,; ~\gamma_{11}=1.00\,, ~\gamma_{22}=25.00\,, ~\gamma_{33}=4.00$\,.}
\label{fig:simpar04}
\end{figure}

Somehow, the resident supports the invasion of its own area. Diffusion and noise enhance the mixing of resident and invaders at the front. 
Therefore, all three together jump on the stable limit cycle of coexistence and invade the remaining invader-free area, cf. 
Figure~\ref{fig:simpar04}.

\FloatBarrier 

The following Figure~\ref{fig:simpar04col} shows that weakly correlated noise still allows for invasion but stronger correlated does not.
\begin{figure}[!ht]
  \centering
\subfigure{\includegraphics[width=0.19\textwidth]{figs_eps/30000All}}
\subfigure[200]{\includegraphics[width=0.19\textwidth]{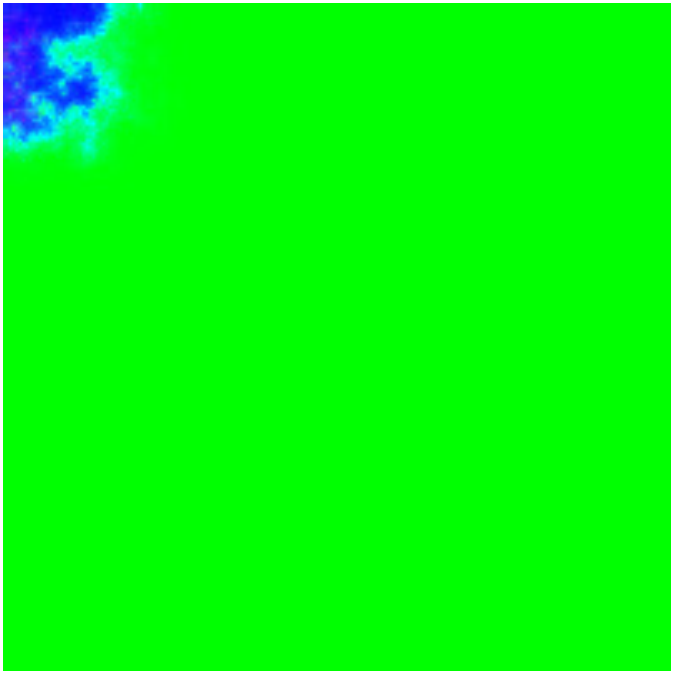}}
\subfigure[500]{\includegraphics[width=0.19\textwidth]{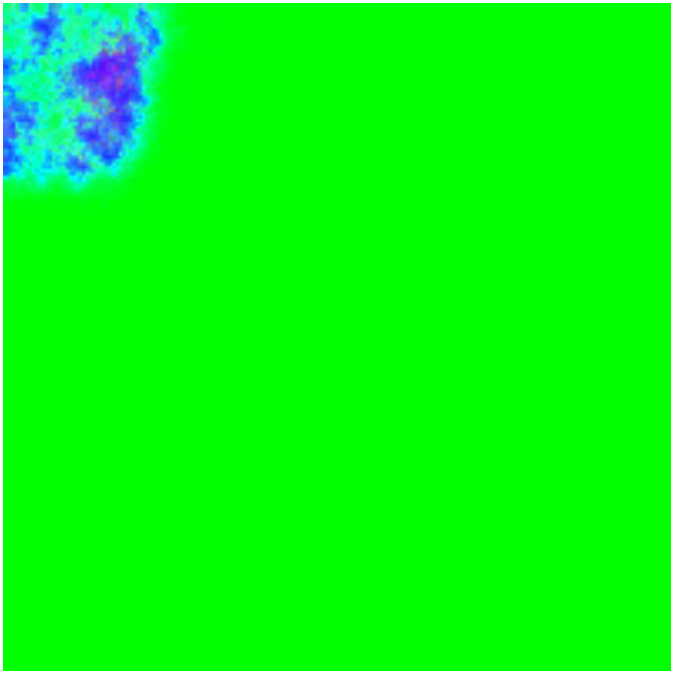}}
\subfigure[2000]{\includegraphics[width=0.19\textwidth]{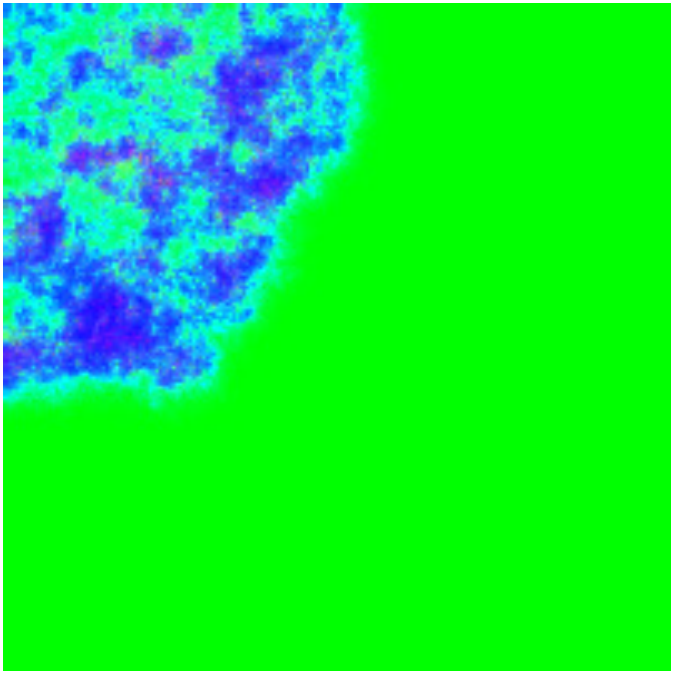}}
\subfigure[4000]{\includegraphics[width=0.19\textwidth]{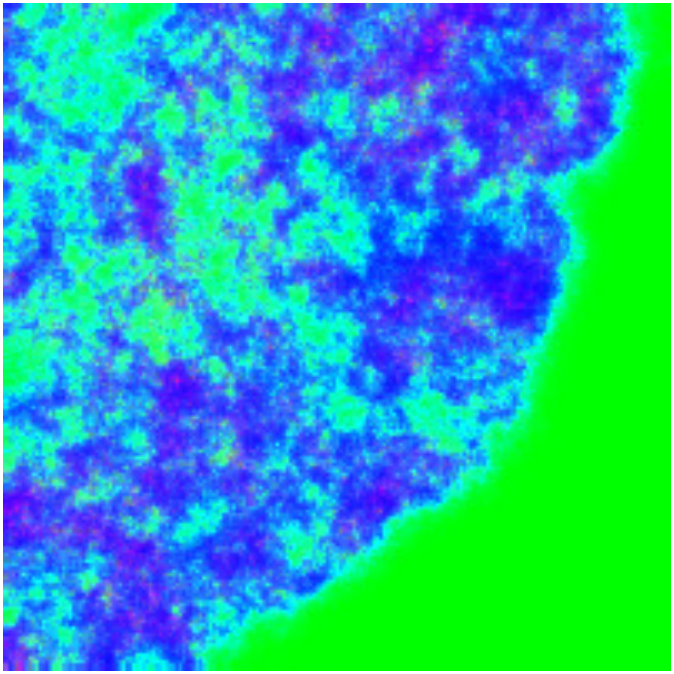}}\\
\subfigure[t=0]{\includegraphics[width=0.19\textwidth]{figs_eps/40000white}}
  \subfigure[100]{\includegraphics[width=0.19\textwidth]{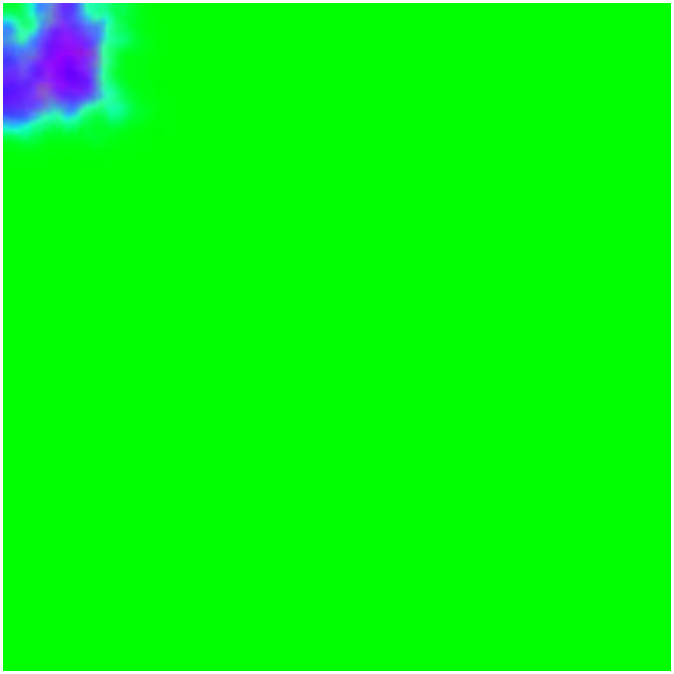}}
\subfigure[150]{\includegraphics[width=0.19\textwidth]{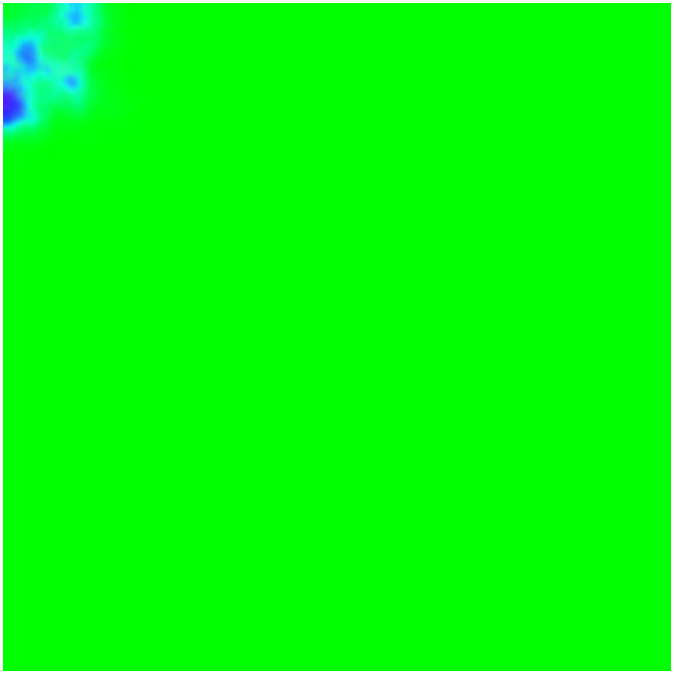}}
\subfigure[500]{\includegraphics[width=0.19\textwidth]{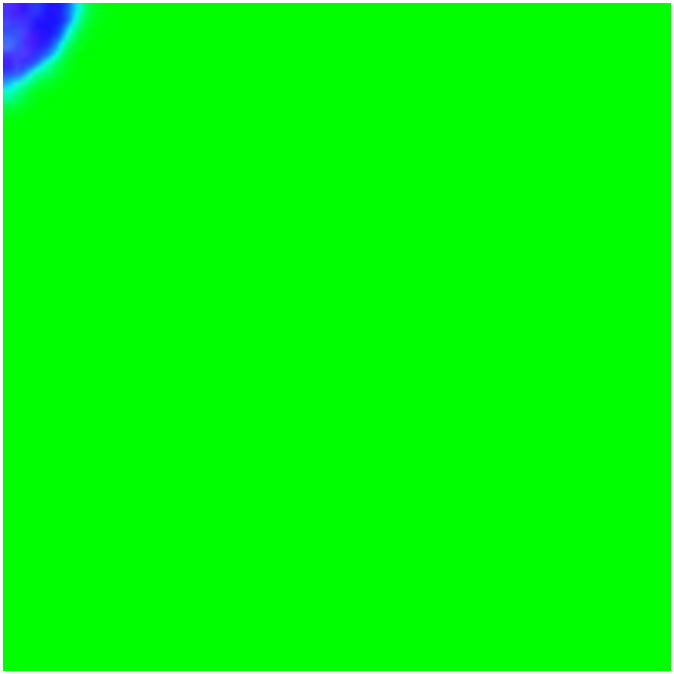}}
\subfigure[1200]{\includegraphics[width=0.19\textwidth]{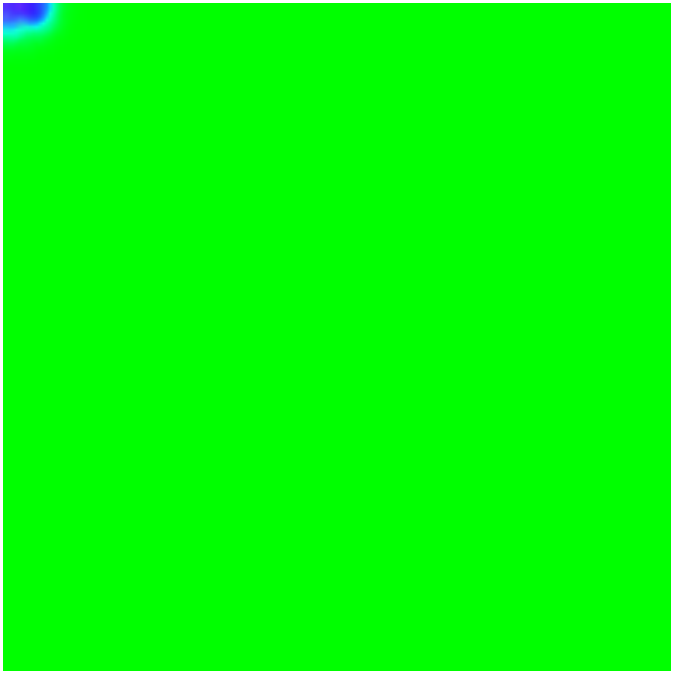}}
  \caption{Same parameters as in Fig.~\ref{fig:simpar04}. The invasion is successful for colored noise and correlation lengths~$\tau=\lambda=1$ but unsuccessful for~$\tau=\lambda=15$. Other parameters~$\epsilon=0.001$, $\omega_{ii}=0.05$.}
  \label{fig:simpar04col}
\end{figure}

\FloatBarrier 

After some difficulties at the beginning that can be the end for the invader at lower noise intensities, the purple invader patch turns into 
the blue of the limit cycle. Finally, the resident survives but has to share its habitat with the aliens. However, if the nonlinear response 
 or colored noise is applied, the invasion can be stopped and rolled back again.


\subsection{Invasions and noise IIb}

Now, it is assumed that the limit cycle of resident and invaders has already invaded most of the area and only a small part is left for the 
resident alone. Again, the parameters (\ref{eq:parabinoise}) are used. For this situation several interesting patterns appear 
that are again purely due to different properties of the environmental noise.

\subsubsection{Dependence on noise intensity}
\label{sec:noisedependence}

\begin{figure}[!ht]
\center
\includegraphics[width=0.75\textwidth]{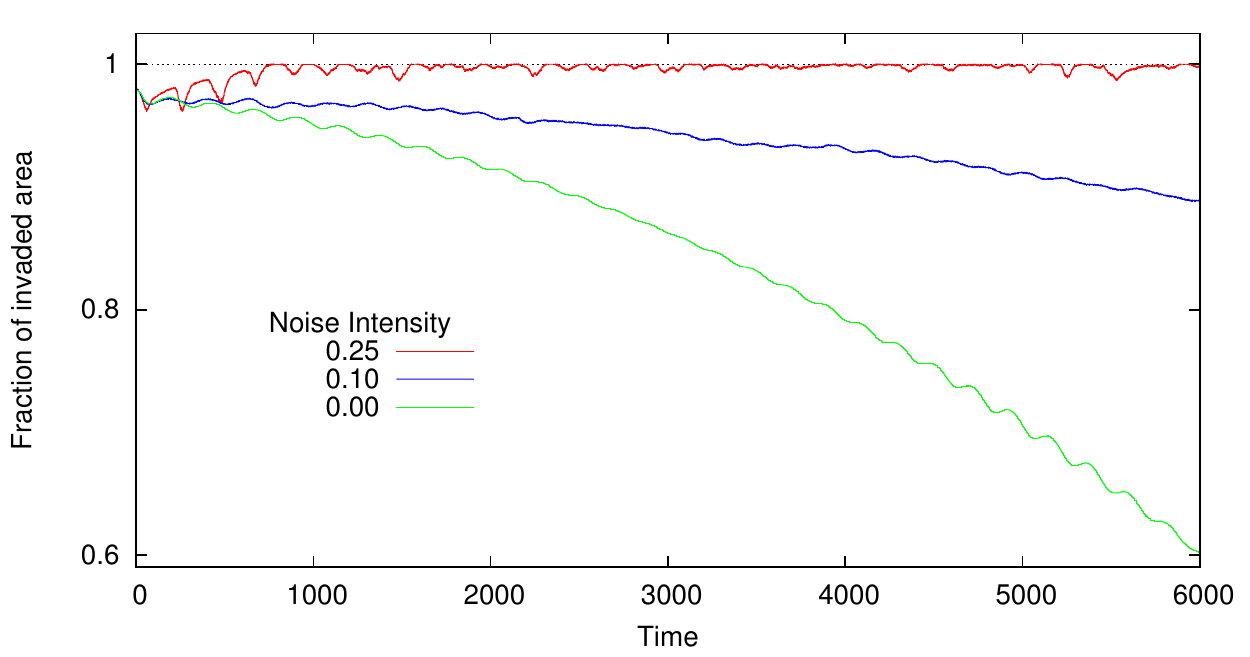}
\caption{Decline resp. growth of invaded area for increasing white noise intensity.}
\label{fig:invareaoscilla2}
\end{figure}

\begin{figure}[!ht]
\center
\subfigure{\includegraphics[width=0.19\textwidth]{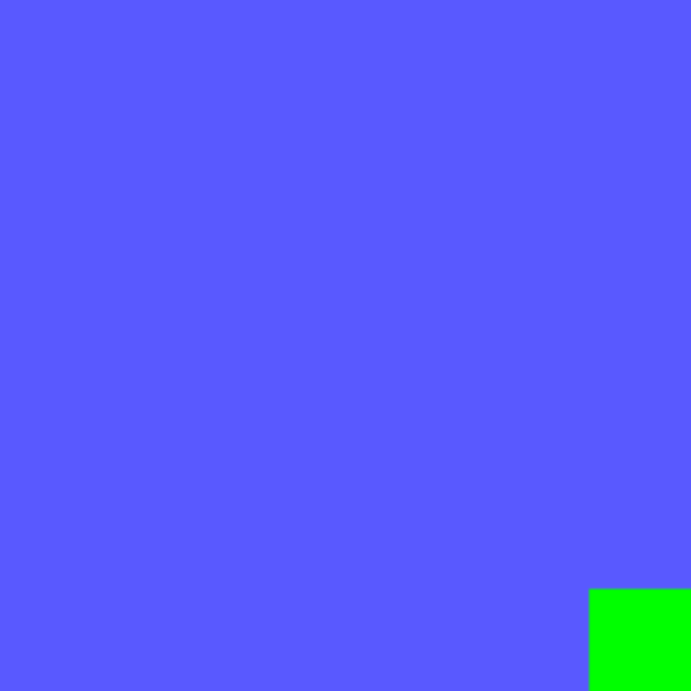}}
\subfigure{\includegraphics[width=0.19\textwidth]{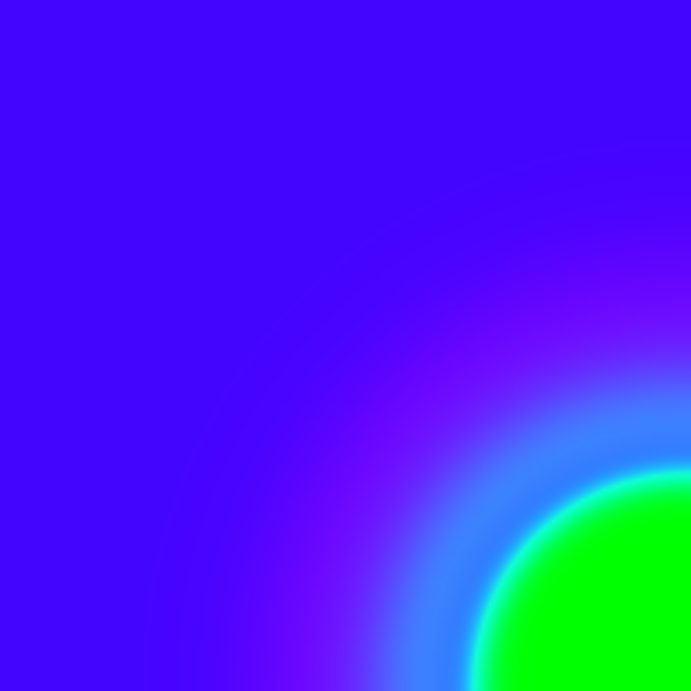}}
\subfigure{\includegraphics[width=0.19\textwidth]{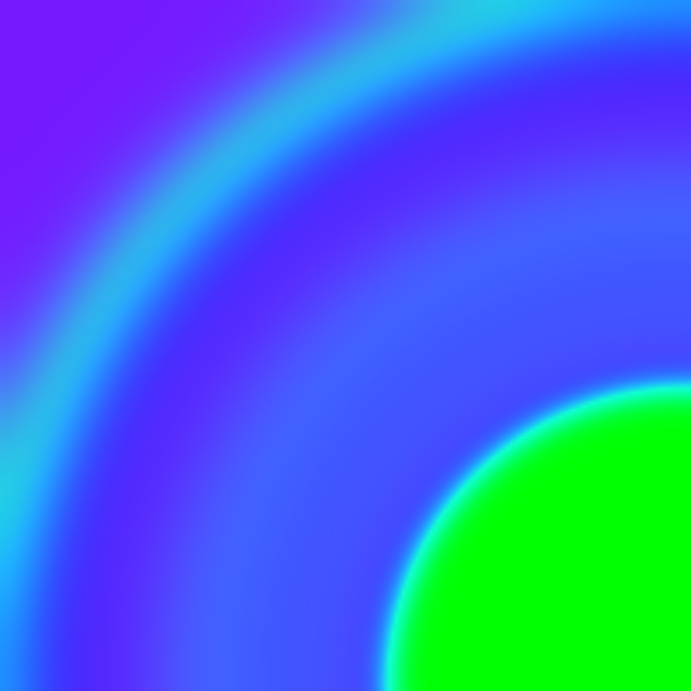}}
\subfigure{\includegraphics[width=0.19\textwidth]{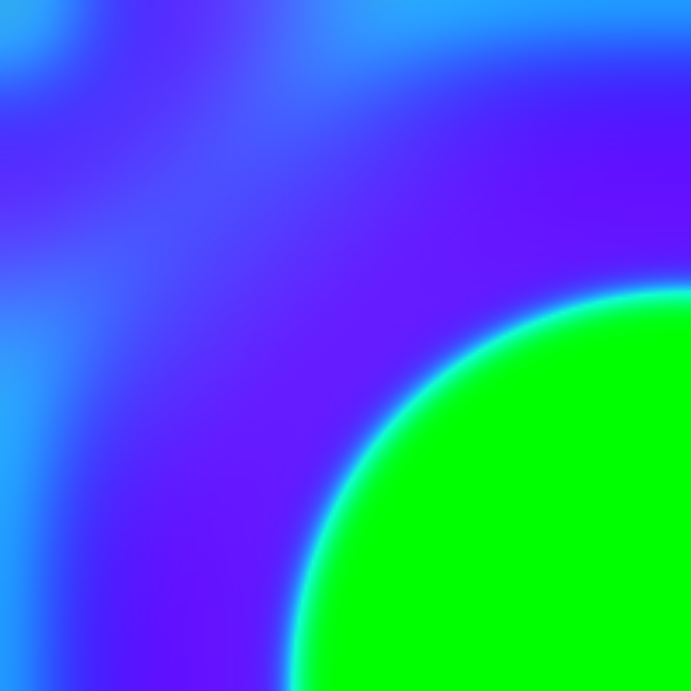}}
\subfigure{\includegraphics[width=0.19\textwidth]{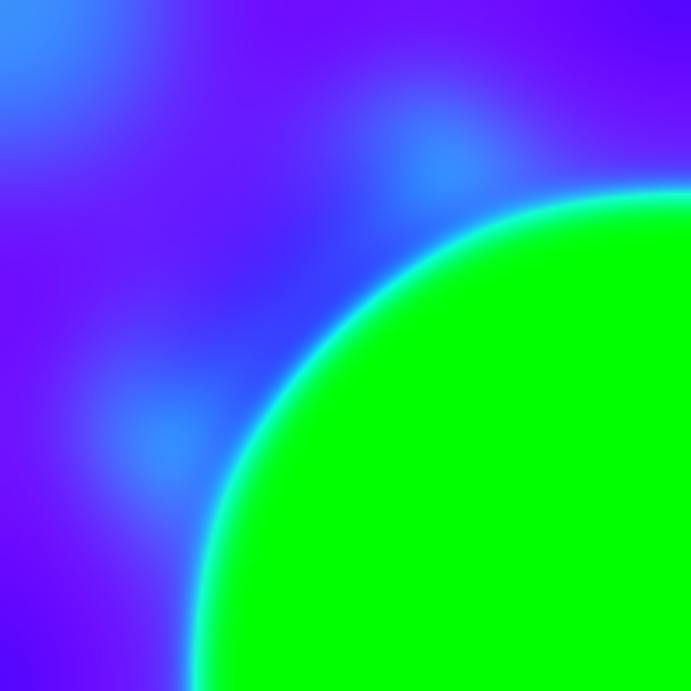}}
\subfigure[t=0]{\includegraphics[width=0.19\textwidth]{figs_eps/40000white}}
\subfigure[1500]{\includegraphics[width=0.19\textwidth]{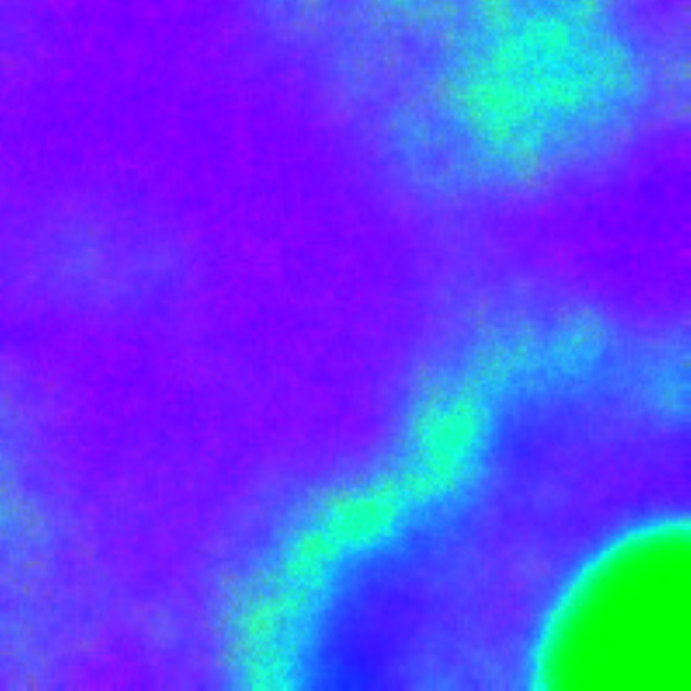}}
\subfigure[3000]{\includegraphics[width=0.19\textwidth]{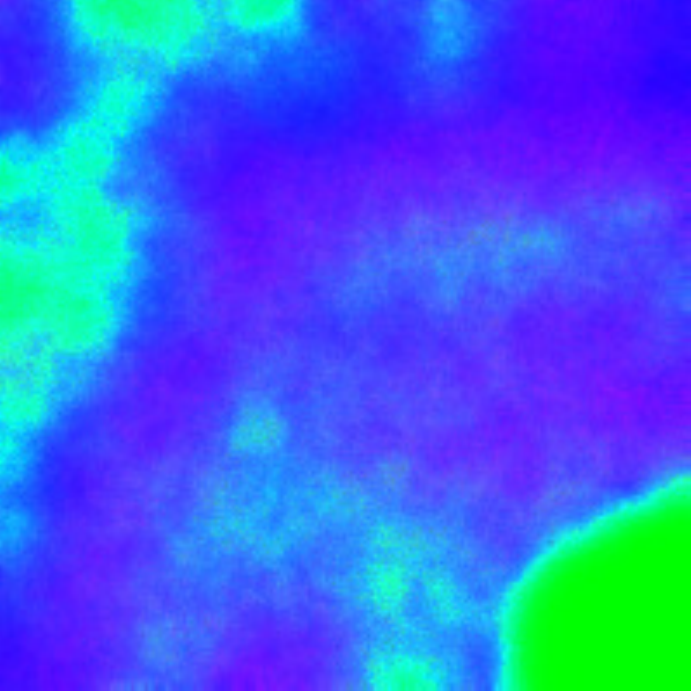}}
\subfigure[4500]{\includegraphics[width=0.19\textwidth]{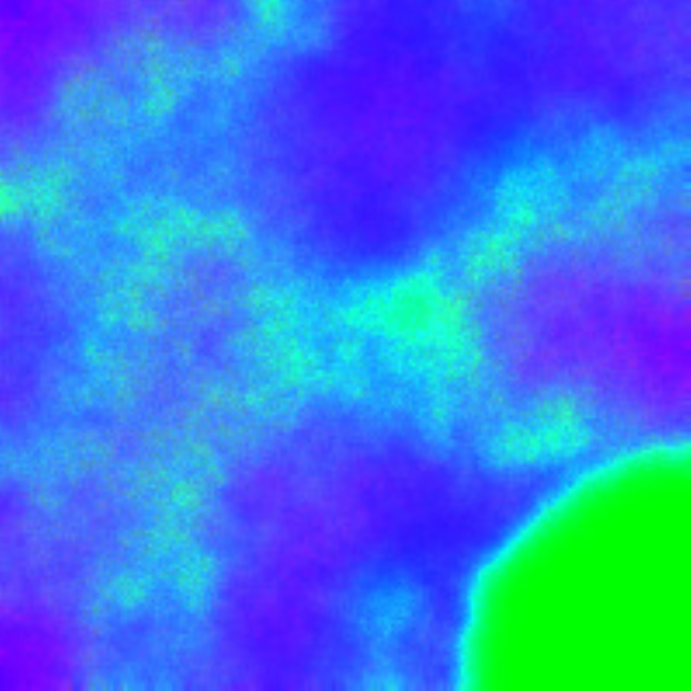}}
\subfigure[6000]{\includegraphics[width=0.19\textwidth]{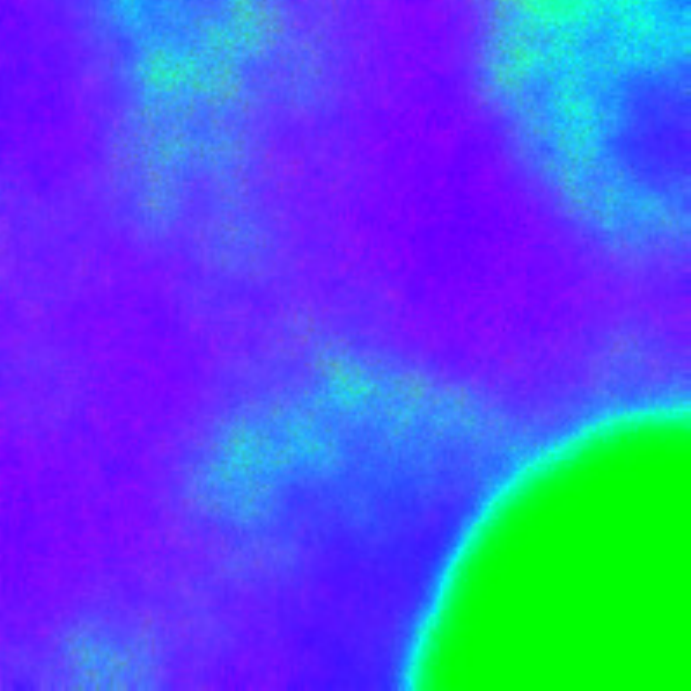}}
\caption{Finally successful defeat of invasion for no noise (upper row) resp. white noise with $\omega_{ii}=0.1\,; ~i=1,2,3$.}
\label{fig:simpar06}
\end{figure}

It is surprising that the resident turns out to be strong enough to defeat the invasion as far the noise is below a subcritical threshold. The fraction of invaded area over time is plotted in Figure\,\ref{fig:invareaoscilla2}.
An example for the defeat of invasion is given in Figure\,\ref{fig:simpar06}.

\FloatBarrier 

\begin{figure}[!ht]
\center
\subfigure[t=100]{\includegraphics[width=0.19\textwidth]{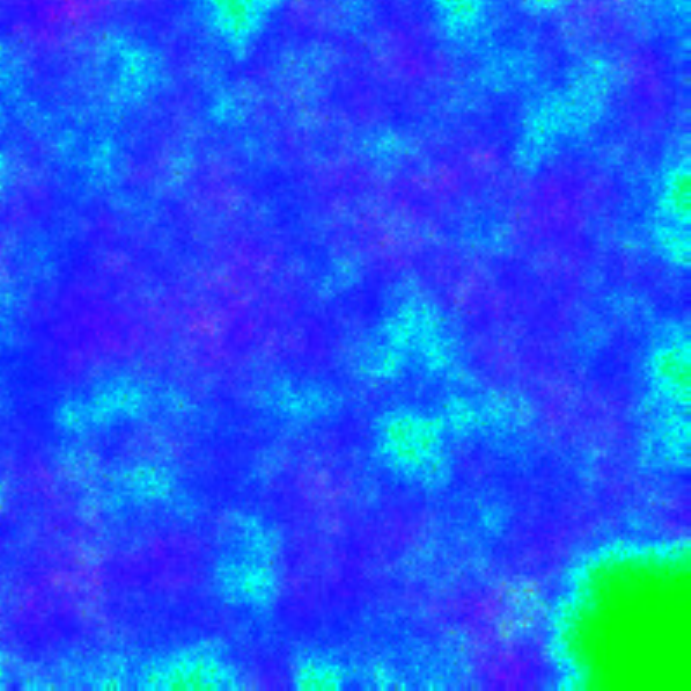}}
\subfigure[1000]{\includegraphics[width=0.19\textwidth]{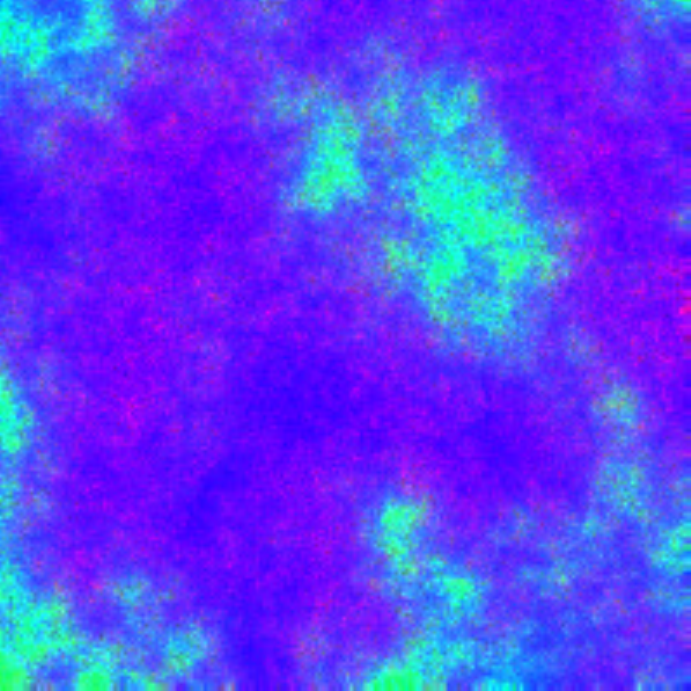}}
\subfigure[2000]{\includegraphics[width=0.19\textwidth]{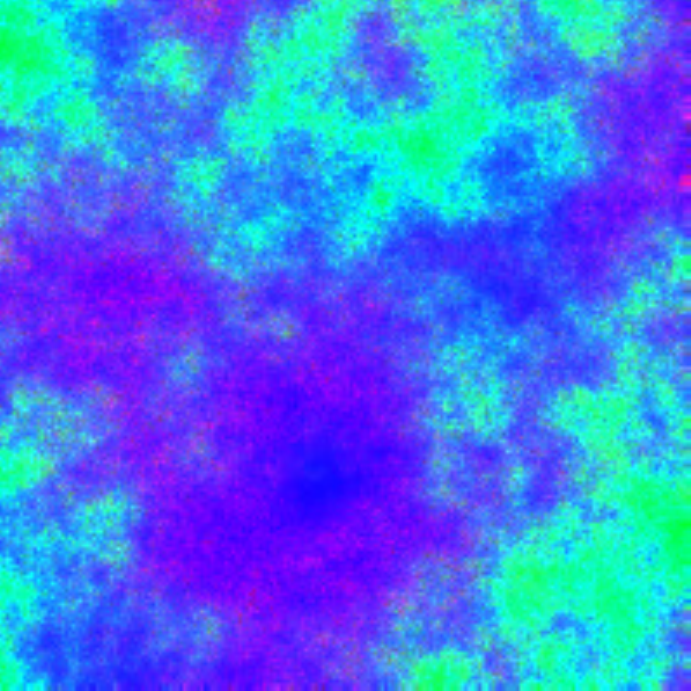}}
\subfigure[4500]{\includegraphics[width=0.19\textwidth]{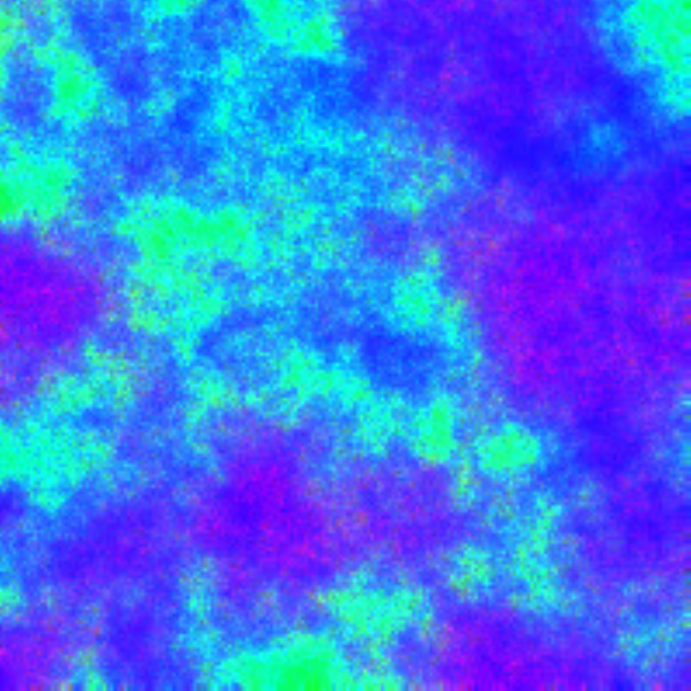}}
\subfigure[6000]{\includegraphics[width=0.19\textwidth]{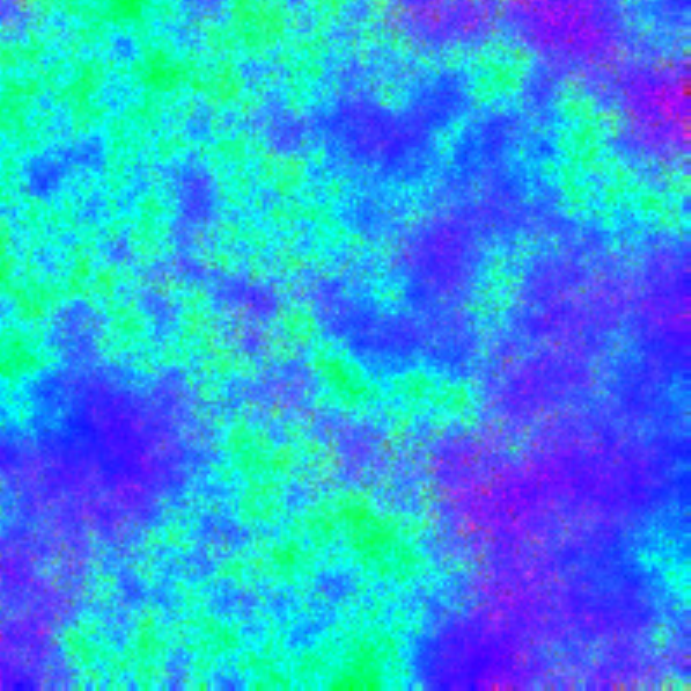}}
\caption{Cloudy invasion for linear white noise with $\omega_{ii}=0.25\,; ~i=1,2,3$. Initial condition as in Figure\,\ref{fig:simpar06}.}
\label{fig:simpar07}
\end{figure}

A preliminary conclusion is that increasing linearly density-dependent white noise supports invasion. In
Figure~\ref{fig:simpar07}, the cloudy result for $\omega_{ii}=0.25$ is shown.\\

\begin{figure}[!ht]
\center
\subfigure[t=50]{\includegraphics[width=0.19\textwidth]{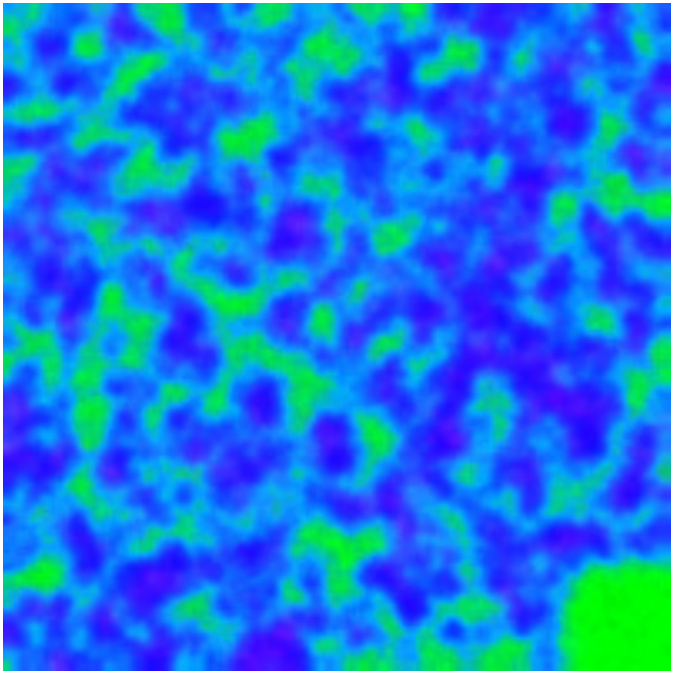}}
\subfigure[500]{\includegraphics[width=0.19\textwidth]{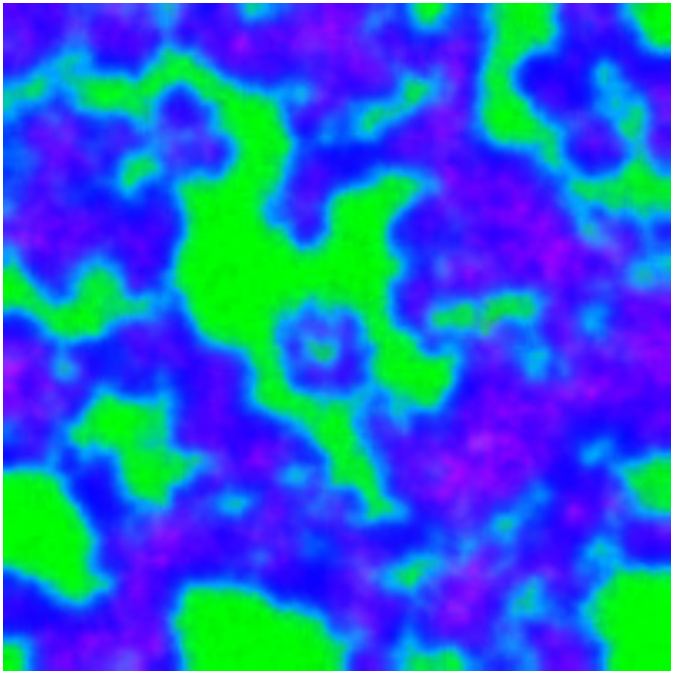}}
\subfigure[2000]{\includegraphics[width=0.19\textwidth]{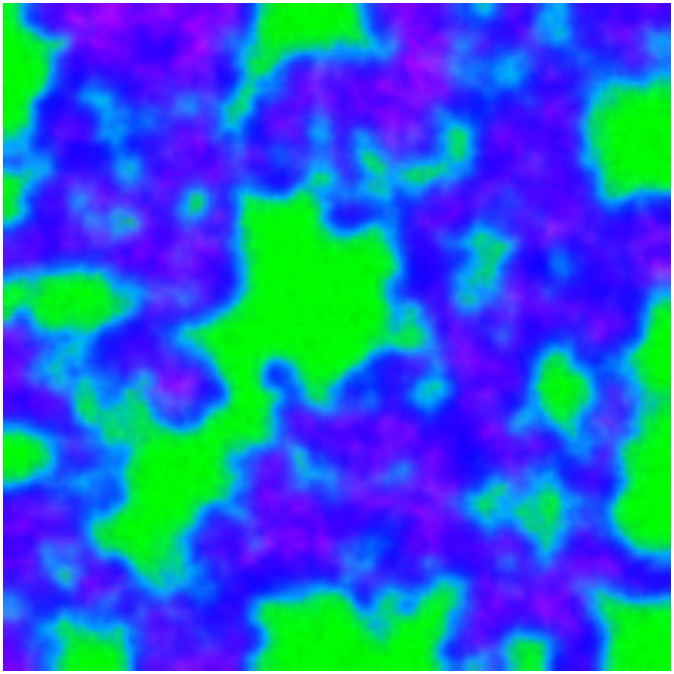}}
\subfigure[7500]{\includegraphics[width=0.19\textwidth]{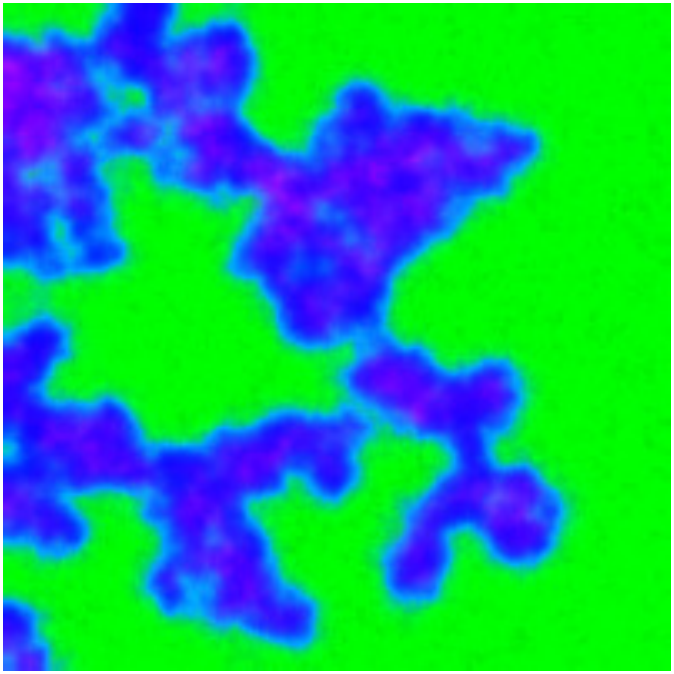}}
\subfigure[10000]{\includegraphics[width=0.19\textwidth]{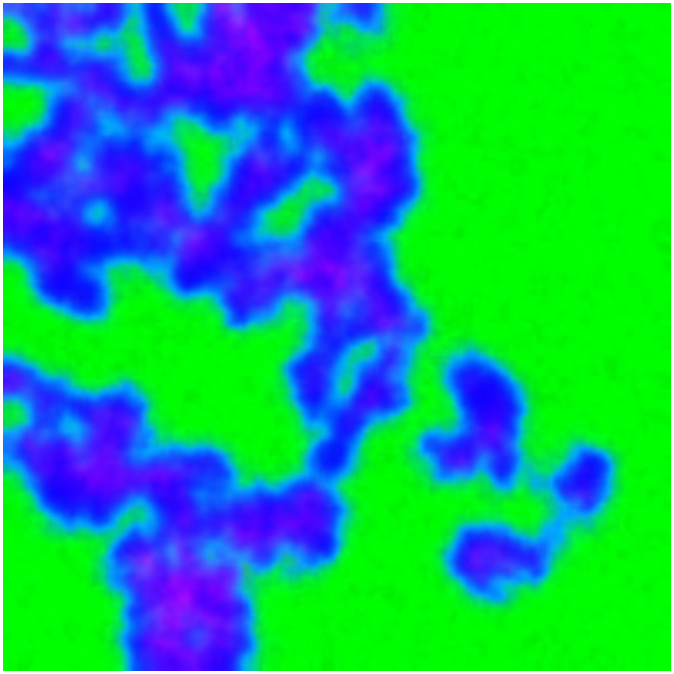}}
\caption{Resident and invader fight under colored noise for territory creating a highly irregular pattern for
  $\omega_{ii}=0.1\,, ~\epsilon=0.001\,, ~\tau=20\,, \lambda=20$\,. Initial condition as in Figure\,\ref{fig:simpar06}.}
\label{fig:spatTempw01tau20lambda20}
\end{figure}

Increasing the noise intensity and both correlation
lengths~($\tau=\lambda=20$) leads to a situation where native and
resident gain and lose control over parts of the spatial domain in an
alternating fashion (Figure~\ref{fig:spatTempw01tau20lambda20}). It
seems that the resident is slowly getting the upper hand: at~$t=7500$
roughly two thirds of the domain are occupied by natives, at~$t=10000$
the resident has lost a few areas and displaced the invader in a few
others.

\FloatBarrier 

\subsubsection{Metapopulation patches}
\label{sec:metapatch}

\begin{figure}[!ht]
\center
\subfigure[t=500]{\includegraphics[width=0.19\textwidth]{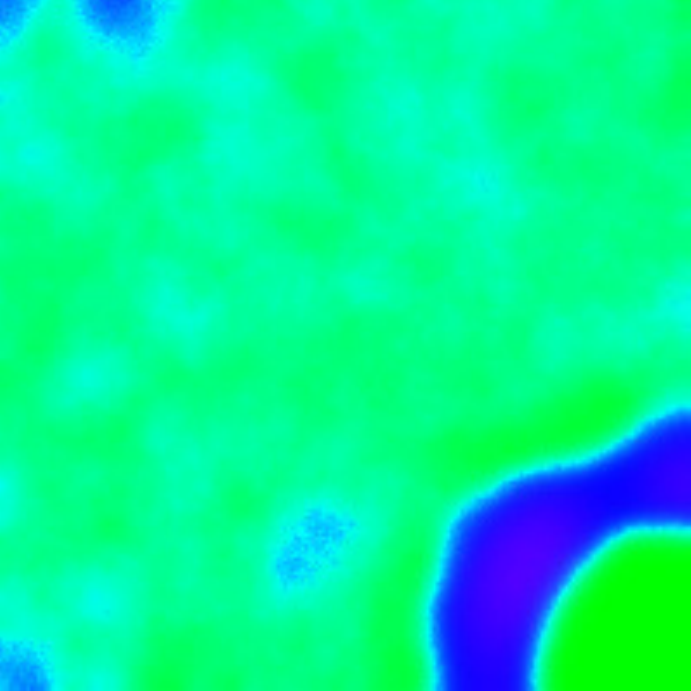}}
\subfigure[1000]{\includegraphics[width=0.19\textwidth]{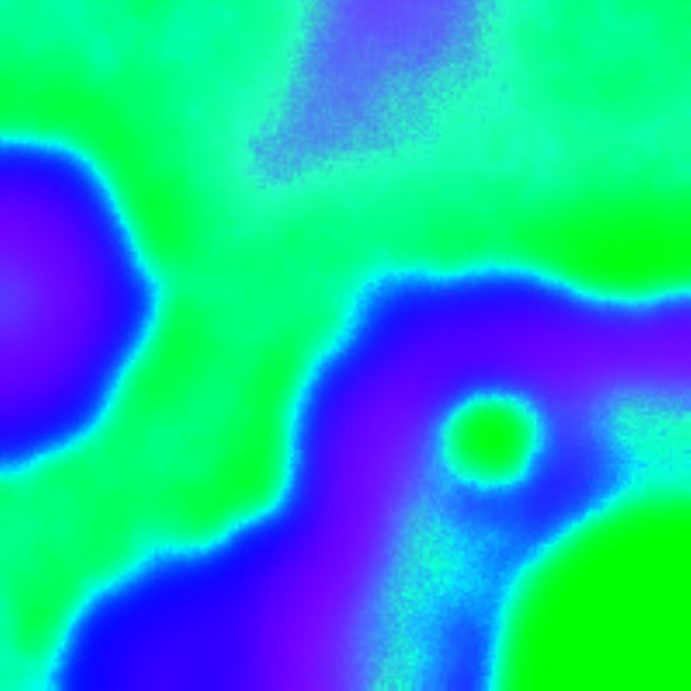}}
\subfigure[1500]{\includegraphics[width=0.19\textwidth]{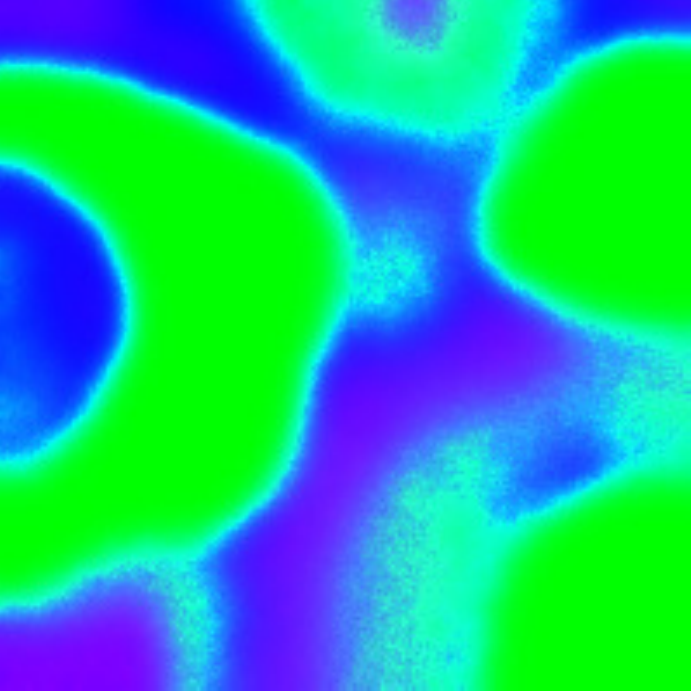}}
\subfigure[2300]{\includegraphics[width=0.19\textwidth]{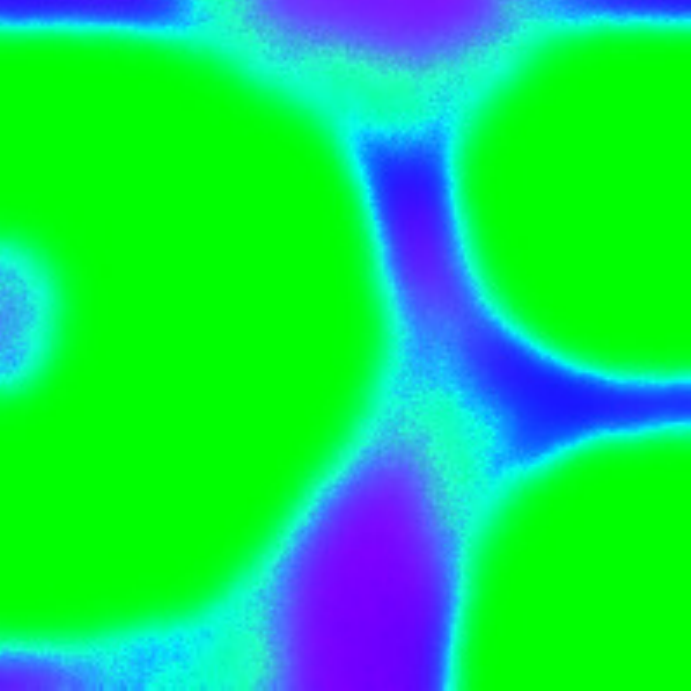}}
\subfigure[6000]{\includegraphics[width=0.19\textwidth]{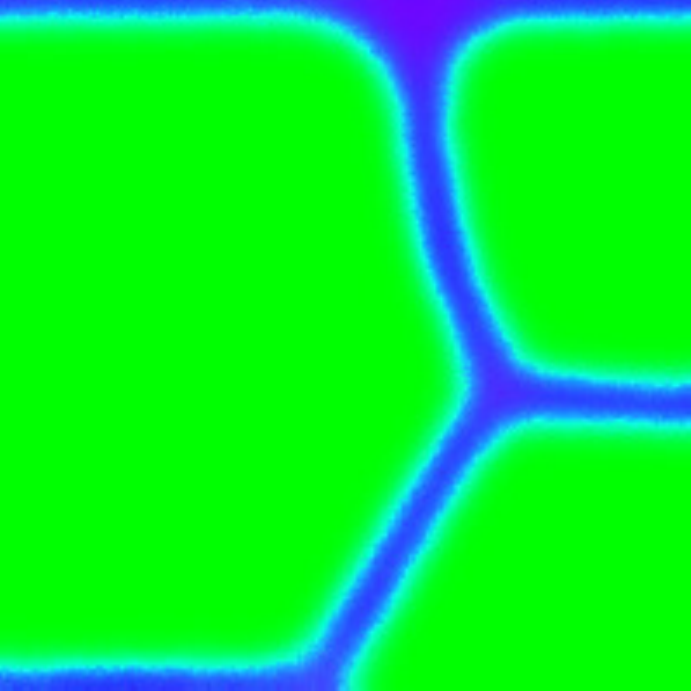}}
\caption{Stable formation of spatially separated patches of the resident population for white noise intensity $\omega_{ii}=0.25\,; ~i=1,2,3\,; ~\gamma_{11}=1.0\,, ~\gamma_{22}=25.0\,,  ~\gamma_{33}=4.0$\,. Initial condition as in Figure\,\ref{fig:simpar06}.}
\label{fig:simpar08}
\end{figure}

This does not necessarily change for nonlinear noise, however, one setting is found where the fraction of invaded area is dropped down from initially 98\% to 21\%. The resident population splits into three metapopulations, spatially separated by the invader populations, cf. Figure\,\ref{fig:simpar08}.

As observed above, higher noise intensities~$\omega$ help the invader
to establish by stabilising the coexistence limit cycle. In contrast,
stronger correlations i.e. increasing correlation lengths~$\tau$ in
time or~$\lambda$ in space generally enable the native species to
displace the invaders. Only for relatively high correlation lengths of
10 or above the native species is able to form patches and avoid a
cloudy invasion as in Figure~\ref{fig:simpar07}.\\

\begin{figure}[!ht]
\center
\subfigure[t=340]{\includegraphics[width=0.19\textwidth]{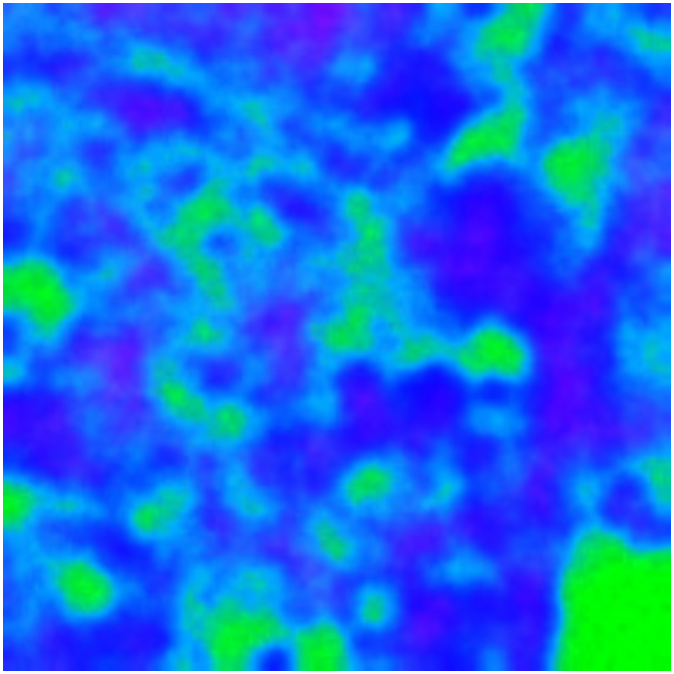}}
\subfigure[1000]{\includegraphics[width=0.19\textwidth]{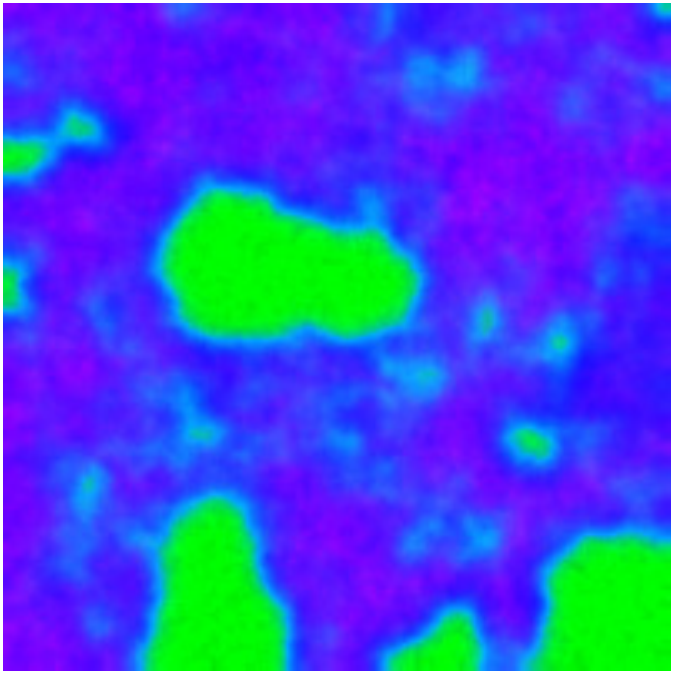}}
\subfigure[2000]{\includegraphics[width=0.19\textwidth]{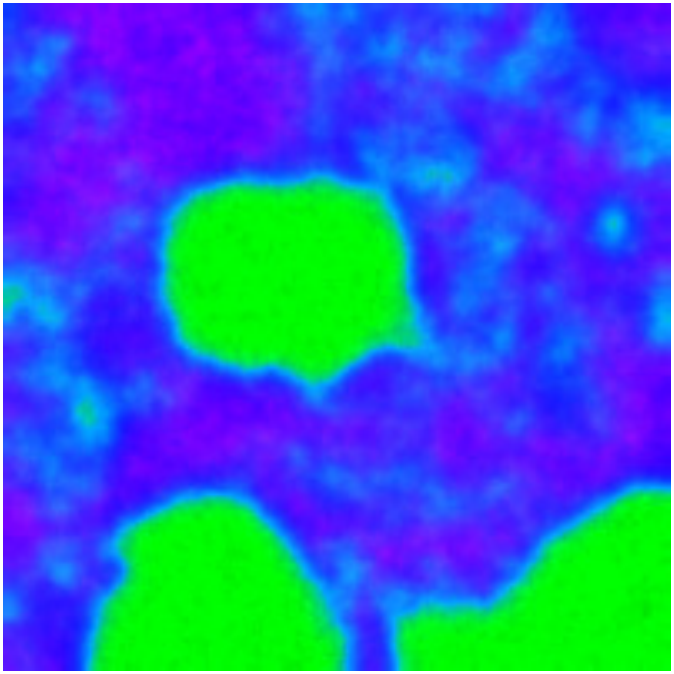}}
\subfigure[3000]{\includegraphics[width=0.19\textwidth]{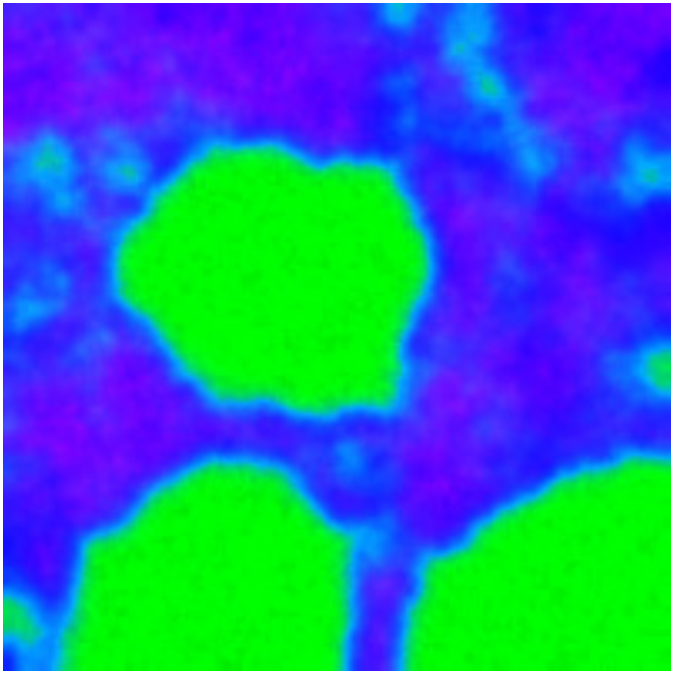}}
\subfigure[6000]{\includegraphics[width=0.19\textwidth]{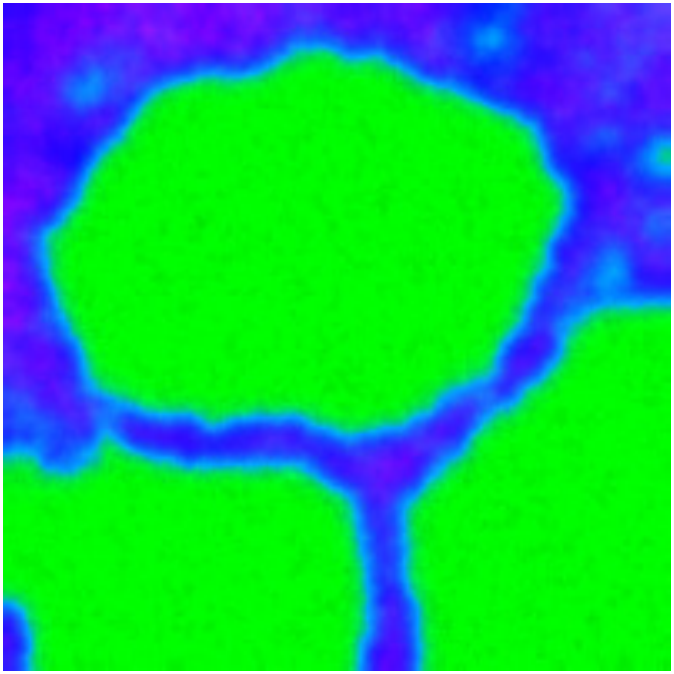}}
\caption{Resident slowly displaces invader while forming seemingly stable spatially separated patches, similar to the results in Figure~\ref{fig:simpar08} but without nonlinearity in the noise term. Parameters: $\omega_{ii}=0.03\,, ~\epsilon=0.001\,, ~\tau=1\,, \lambda=15$\,. Initial condition as in Figure\,\ref{fig:simpar06}.}
\label{fig:spatTempw003tau1lambda15}
\end{figure}

In Figure~\ref{fig:spatTempw003tau1lambda15}, for a low noise
intensity~$\omega_{ii}=0.03$, $\tau=1$ and~$\lambda=15$ we observe the
emergence of a quasi-stationary pattern similar to
Figure~\ref{fig:simpar08}.\\

\FloatBarrier 

\subsubsection{Spiral waves}
\label{sec:spiralwaves}

\begin{figure}[!ht]
\center
\subfigure[t=500]{\includegraphics[width=0.19\textwidth]{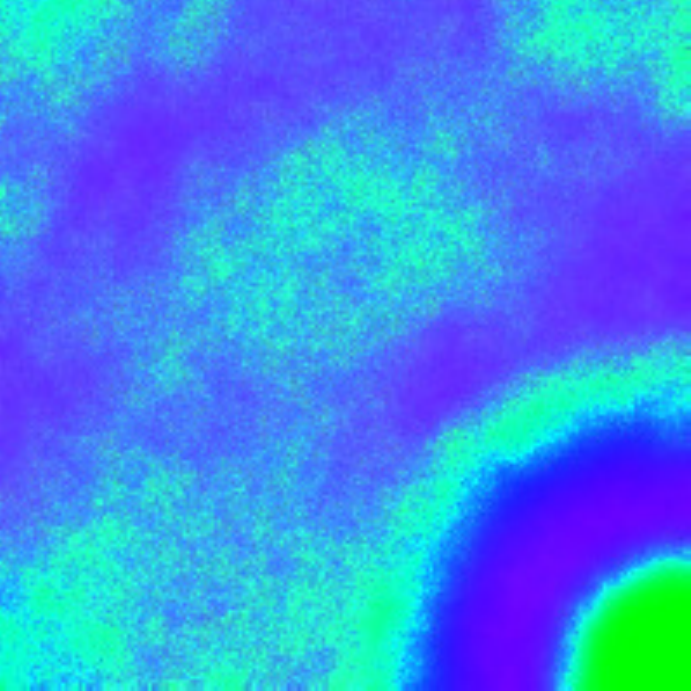}}
\subfigure[1700]{\includegraphics[width=0.19\textwidth]{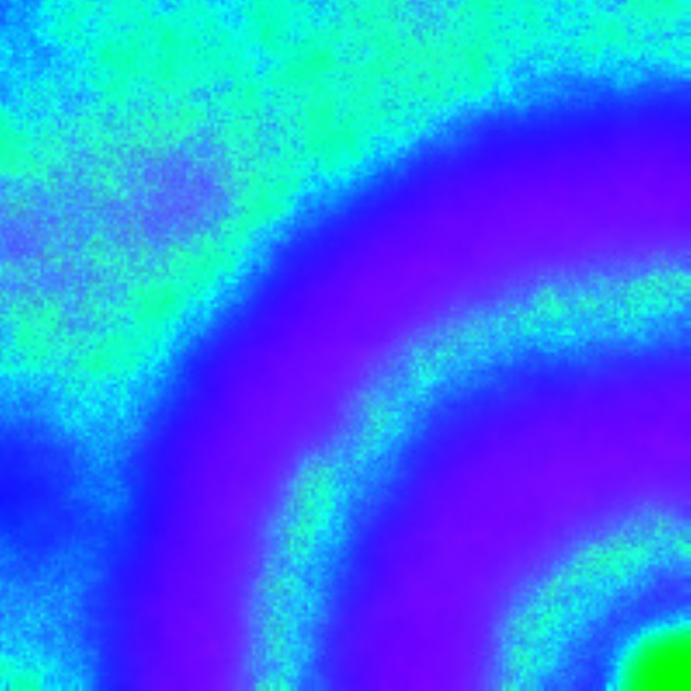}}
\subfigure[2250]{\includegraphics[width=0.19\textwidth]{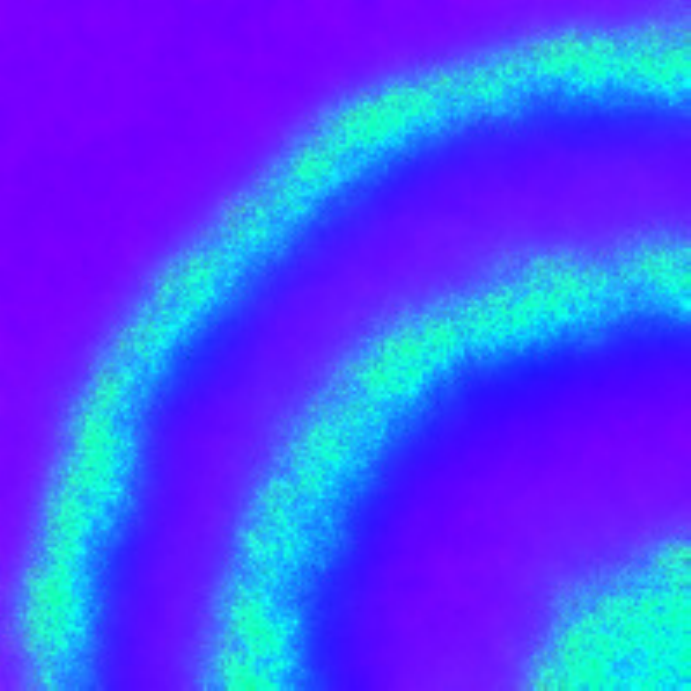}}
\subfigure[5500]{\includegraphics[width=0.19\textwidth]{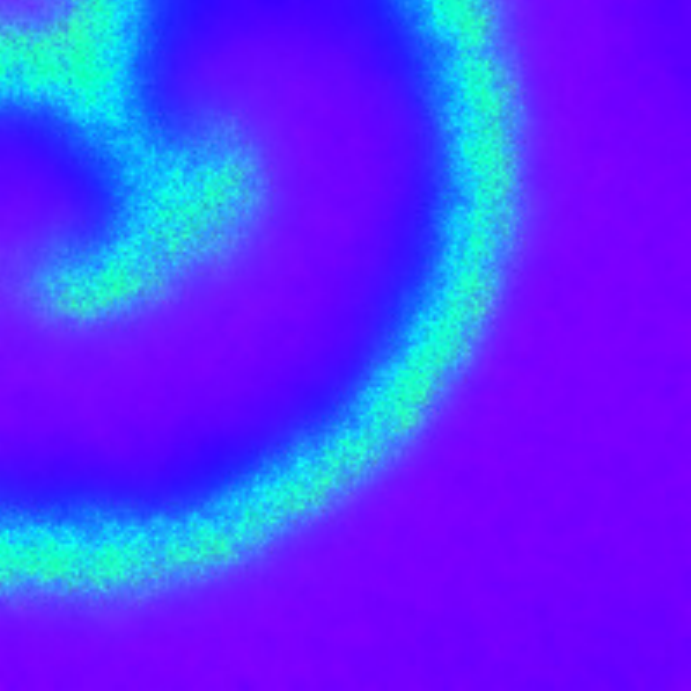}}
\subfigure[6000]{\includegraphics[width=0.19\textwidth]{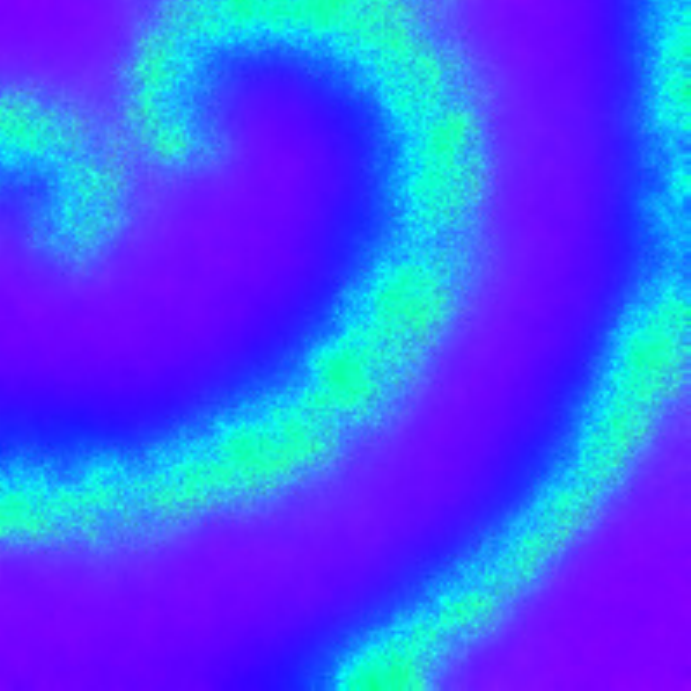}}
\caption{Stable formation of spatially separated patches of the resident population for linear white noise intensity $\omega_{ii}=0.3\,; ~i=1,2,3\,; ~\gamma_{11}=0.5625\,, ~\gamma_{22}=4.0\,,  ~\gamma_{33}=9.0$\,. Initial condition as in Figure\,\ref{fig:simpar06}.}
\label{fig:simpar09}
\end{figure}

Wavy structures are found as well, however, at the cost of full invasion. One example is presented in Figure\,\ref{fig:simpar09}.

\begin{figure}[!ht]
\center
\subfigure[t=1000]{\includegraphics[width=0.19\textwidth]{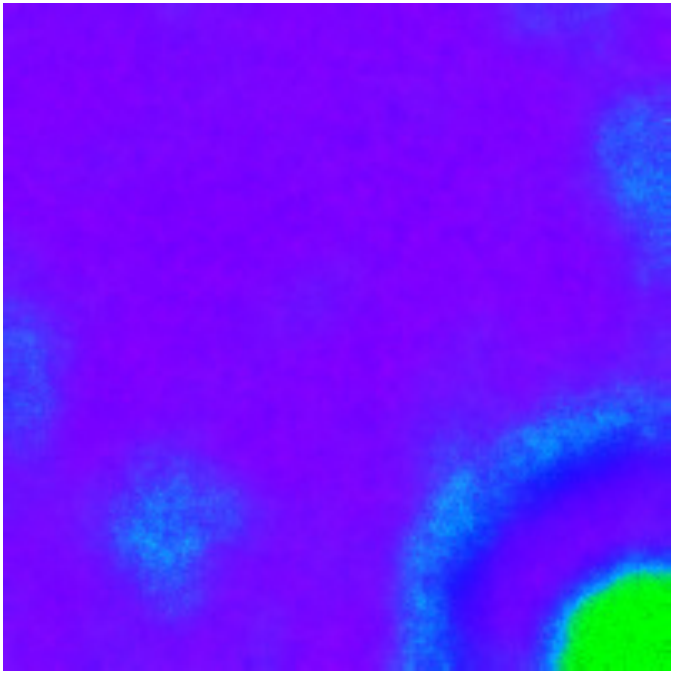}}
\subfigure[t=2000]{\includegraphics[width=0.19\textwidth]{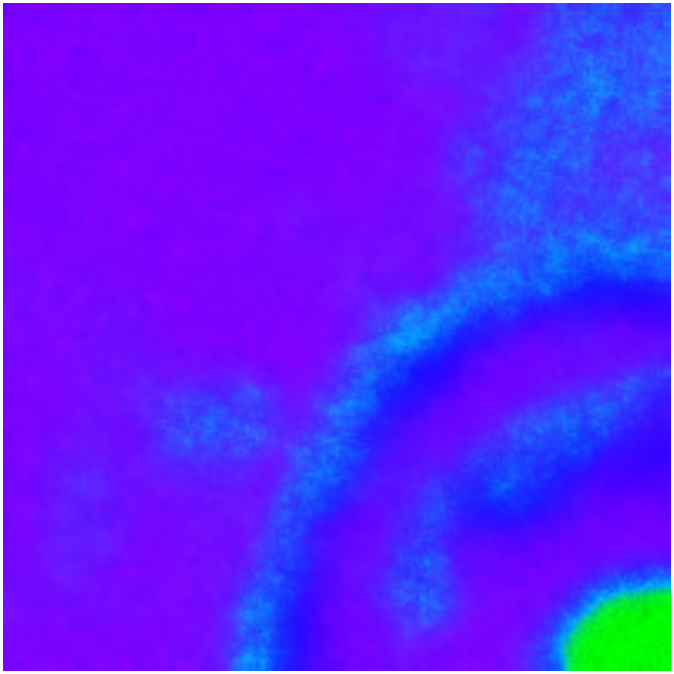}}
\subfigure[t=5000]{\includegraphics[width=0.19\textwidth]{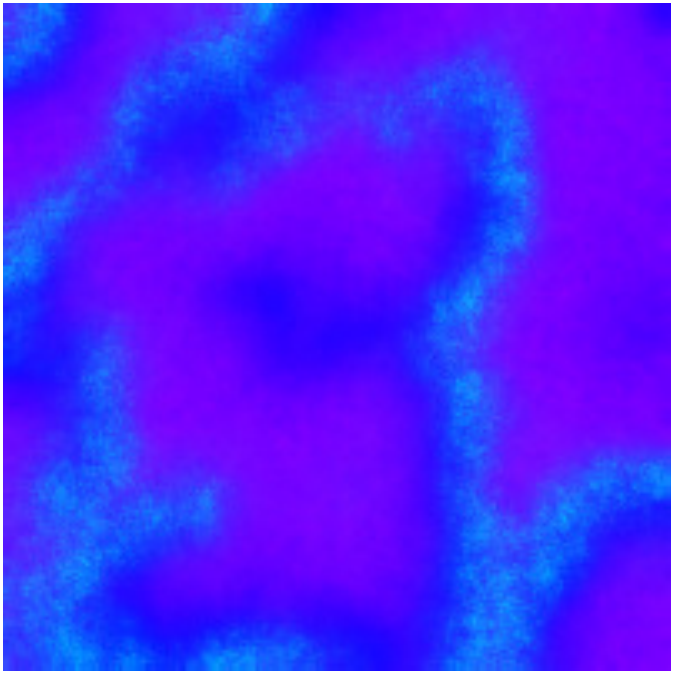}}
\subfigure[t=7000]{\includegraphics[width=0.19\textwidth]{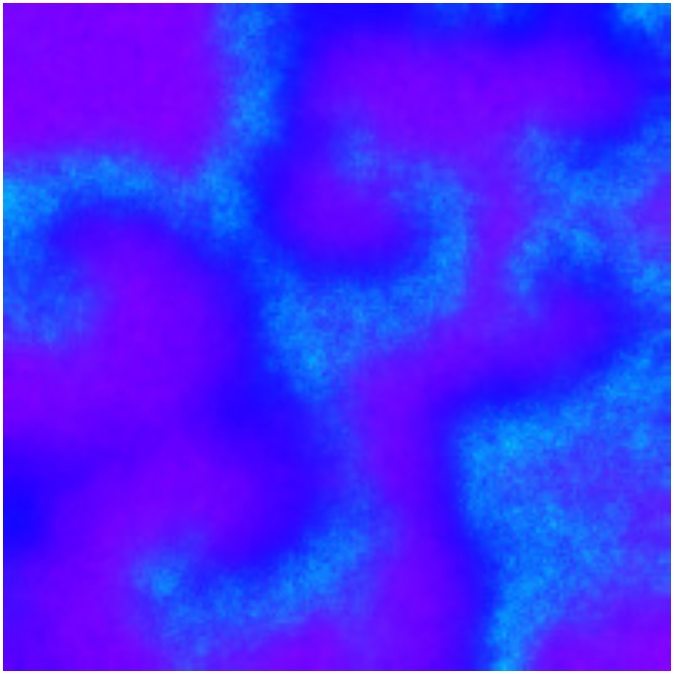}}
\subfigure[t=9000]{\includegraphics[width=0.19\textwidth]{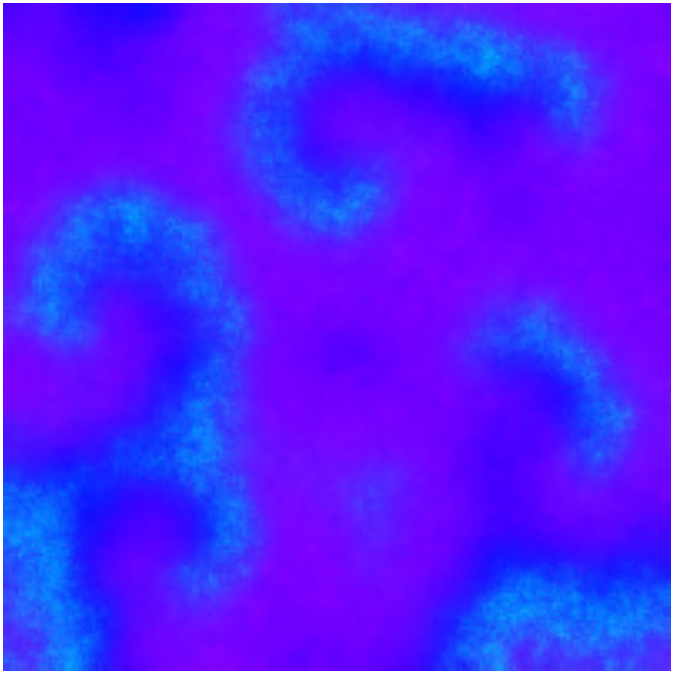}}

\caption{Formation of spiral waves -- similar to Figure~\ref{fig:simpar09} but less regular -- for colored noise intensity
  $\omega_{ii}=0.03\,; ~i=1,2,3\,; ~\epsilon=0.001\,, ~\tau=1\,, ~\lambda=1\,, ~\gamma_{11}=0.5625\,, ~\gamma_{22}=4.0\,,
  ~\gamma_{33}=9.0$\,.
  Initial condition as in Figure\,\ref{fig:simpar06}.}
\label{fig:spatTempsaturatedNoisew003tau1lambda1spiral1}
\end{figure}

For nonlinear noise, the spiral waves seen in
Figure~\ref{fig:simpar09} can also be observed if the noise is
colored. For small values of spatial and temporal correlation
lengths~($\tau=\lambda=1$), smaller and slightly more irregular spiral
waves can be observed, cf. Figure~\ref{fig:spatTempsaturatedNoisew003tau1lambda1spiral1}.\\

\begin{figure}[!ht]
\center
\subfigure[t=1000]{\includegraphics[width=0.19\textwidth]{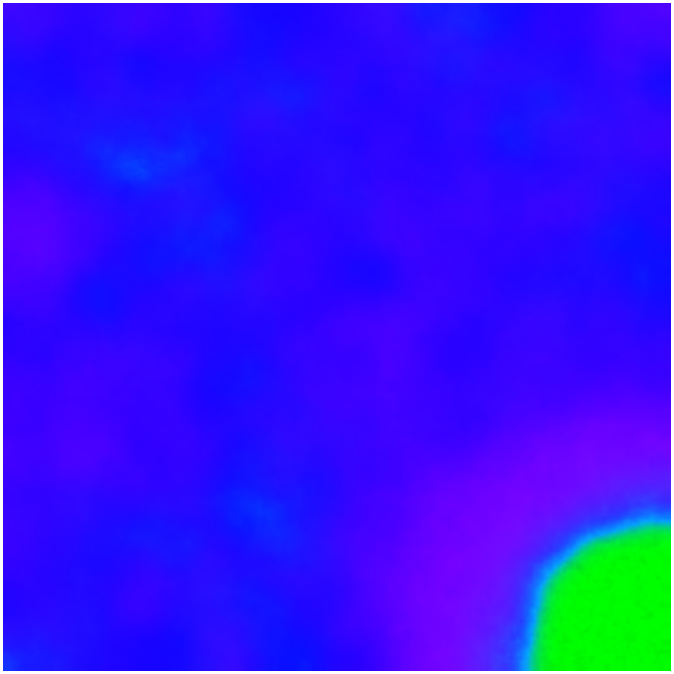}}
\subfigure[t=2000]{\includegraphics[width=0.19\textwidth]{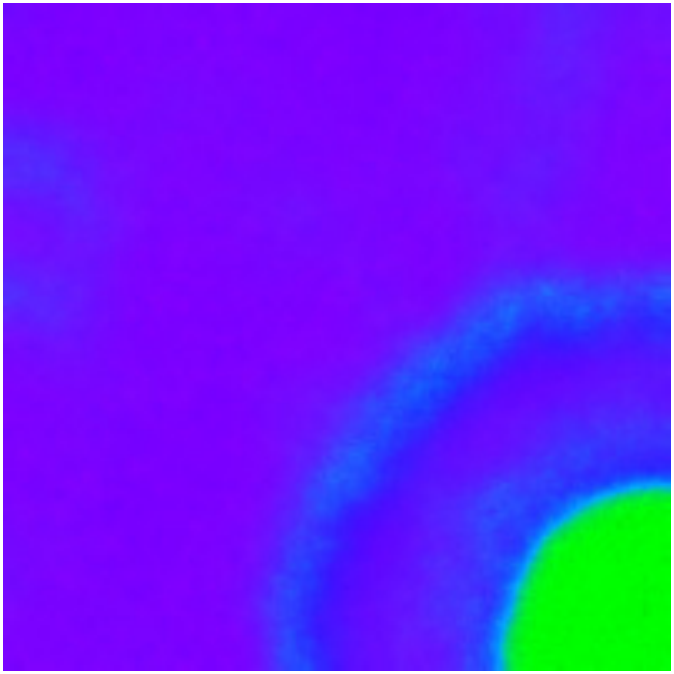}}
\subfigure[t=5000]{\includegraphics[width=0.19\textwidth]{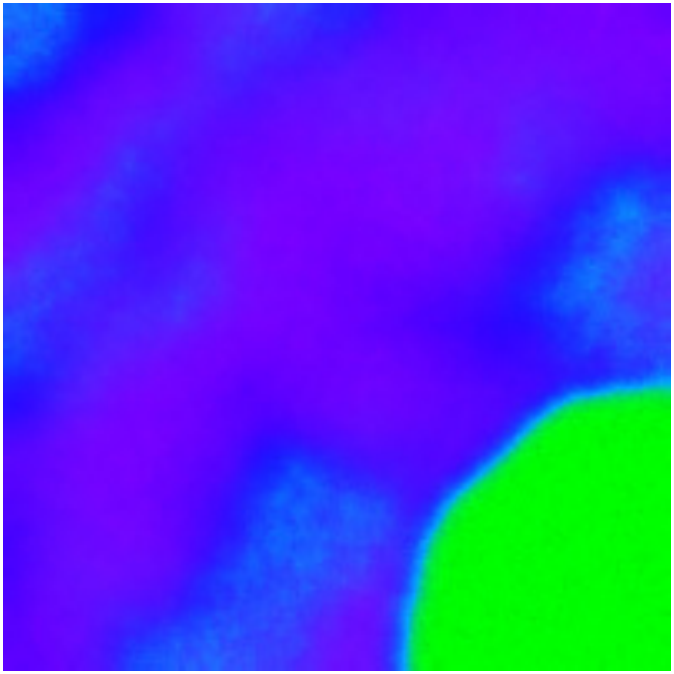}}
\subfigure[t=7000]{\includegraphics[width=0.19\textwidth]{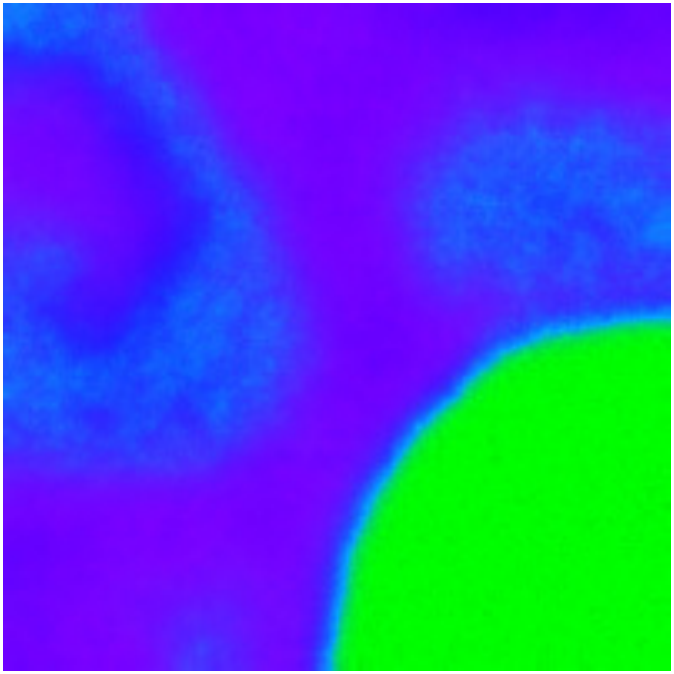}}
\subfigure[t=9000]{\includegraphics[width=0.19\textwidth]{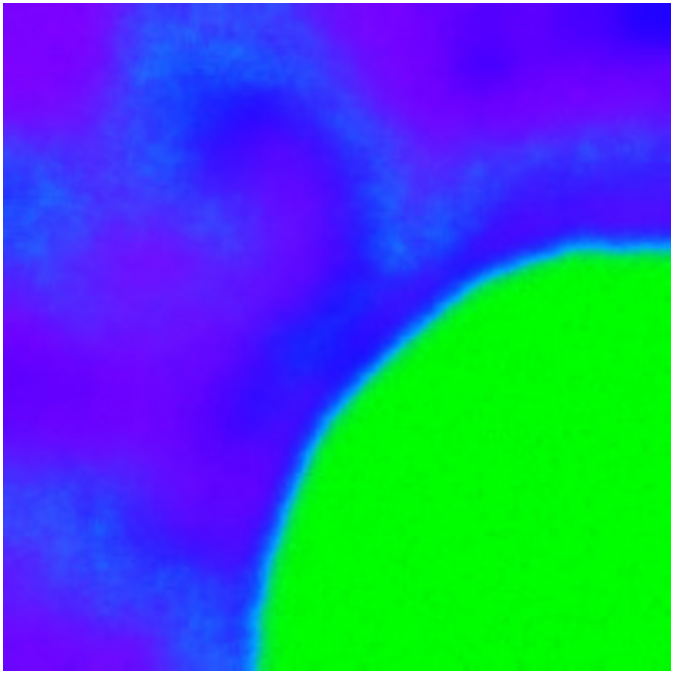}}

\caption{For increased spatial correlation the spiral waves are displaced by a front of natives. Parameters:
  $\omega_{ii}=0.03\,; ~i=1,2,3\,; ~\epsilon=0.001\,, ~\tau=1\,, ~\lambda=5, ~\gamma_{11}=0.5625\,, ~\gamma_{22}=4.0\,, ~\gamma_{33}=9.0$\,.
  Initial condition as in Figure\,\ref{fig:simpar06}.}
\label{fig:spatTempsaturatedNoisew003tau1lambda5spiral2}
\end{figure}

For increased correlation in time or space (e.g. $\tau=1$ as above but
$\lambda=5$), the native population can still maintain a spreading
front that eventually displaces the spiral waves formed in its wake, cf.
Figure~\ref{fig:spatTempsaturatedNoisew003tau1lambda5spiral2}.

\section{Conclusions}

  Populations are exposed to fluctuations of many environmental
  parameters such as nutrient availability, temperature etc. that may
  have positive or adverse effects. Whereas it would be impractical to
  explicitly consider a host of factors that each on their own may
  only have a small influence on growth or decline of the population
  it is possible to represent the collective effect of these factors
  as stochastic environmental variability. The standard model for
  stochastic environmental variability in population dynamics are
  stochastic differential equations with a multiplicative noise term.\\

  In this model, the matrix of maximum noise
  intensities~$\mathbf{\omega}$ is the
  only parameter that can be used for capturing all aspects of
  environmental fluctuations. The standard assumption that
  environmental noise is temporally and spatially uncorrelated
  neglects the fact that many environmental factors are, in fact,
  typically correlated.\\

  Whereas at first glance it seems that this can be convincingly
  justified by assuming that the spatial and temporal correlation
  lengths~$\tau$ and~$\lambda$ of the noise are much shorter than the
  spatiotemporal scale under consideration, we have demonstrated in
  this study that using correlated instead of uncorrelated noise may
  lead to qualitatively very different model behaviour. Thus,
  neglecting possible correlations may, in fact, lead to different
  explanations of the observed system behaviour.\\

  Another implicit assumption that has previously been unquestioned is
  the linear increase of noise intensity with population number. This
  model implies that environmental effects on each individual in a
  population simply add up to an overall effect on the population. In
  our opinion it is likely that for increasing population numbers,
  perturbations should not independently affect each individual but
  rather saturate due to interactions of the individuals so that the
  collective response of the population saturates to a maximum noise
  intensity for large population numbers. This model requires an
  additional matrix~$\mathbf{\gamma}$ which
  characterises the ability of the population to ``buffer'' stochastic
  fluctuations: for low values of~$\gamma_{ii}$, population~$X_i$ is
  exposed to intensities close to the maximum noise level~$\omega_{ii}$
  for low or moderate population numbers whereas for a population with
  a large~$\gamma_{ii}$ the noise intensity~$\omega_{ii}$ is only reached
  for high population numbers.\\
  
  The results presented in this paper have been obtained for purely diagonal noise intensity matrices. More complex forms are left to future work.


\subsection*{Acknowledgements}

H.M. did most of the work on this paper during a semi-sabbatical in Brazil, France and Japan. He acknowledges substantial financial support by the Coordena\c{c}\~ao de Aperfei\c{c}oamento de Pessoal de N\'ivel Superior (CAPES, Brazil), the Funda\c{c}\~ao de Amparo \`a Pesquisa do Estado do Rio Grande do Sul (FAPERGS, Brazil), the Initiative d'Excellence de l'Universit\'e de Bordeaux (IdEX Bordeaux, France), the Japan Society for the Promotion of Science (JSPS, Japan) and last but not least by the Deutscher Akademischer Austauschdienst (DAAD, Germany). And he very much appreciated the scientific expertise and warm hospitality of the host working groups at Santa Maria RS, Bordeaux and Osaka.

\bibliography{refer}
\bibliographystyle{MMNP}

\end{document}